\documentclass[usenatbib]{mnras}
\usepackage{graphicx, amssymb, aas_macros, natbib, bm}
\usepackage{ marvosym }
\usepackage{enumitem}
\usepackage{fancybox}
\usepackage{lipsum}  
\usepackage{gensymb}
\usepackage{csquotes}
\usepackage{xcolor}
\usepackage{amsmath}
\usepackage{multicol}
\usepackage{multirow}
\usepackage{orcidlink}

\graphicspath{{plots/}}

\newcommand{\mtwo}{M_{\rm{200}}}
\newcommand{\rtwo}{R_{\rm{200}}}

\newcommand{\mhalo}{{M}_{\rm{halo}}}

\newcommand{\mstar}{{M}_{\star}}
\newcommand{\minfall}{{M}_{\star, \rm{infall}}}

\newcommand{\msun}{{\rm M}_{\odot}}

\newcommand{\lt}{<}

\DeclareRobustCommand{\ion}[2]{%
\relax\ifmmode
\ifx\testbx\f@series
{\mathbf{#1\,\mathsc{#2}}}\else
{\mathrm{#1\,\mathsc{#2}}}\fi
\else\textup{#1\,{\mdseries\textsc{#2}}}%
\fi}

\title[Quenching Timescales of GOGREEN Satellites]{The GOGREEN Survey: Constraining the Satellite Quenching Timescale in Massive Clusters at \boldmath$z \gtrsim 1$}

\author[Baxter et al.]
{Devontae C. Baxter$^1$\thanks{$\!\!$e-mail: dbaxter@uci.edu}\thanks{$\!\!$LSSTC DSFP Fellow}\orcidlink{0000-0002-8209-2783},
M. C. Cooper$^1$\orcidlink{0000-0003-1371-6019},
Michael L. Balogh$^{2,3}$\orcidlink{0000-0003-4849-9536},
\newauthor
Timothy Carleton$^4$\orcidlink{0000-0001-6650-2853},
Pierluigi Cerulo$^5$\orcidlink{0000-0003-0703-3123},
Gabriella De Lucia$^6$\orcidlink{0000-0002-6220-9104},
\newauthor
Ricardo Demarco$^7$\orcidlink{0000-0003-3921-2177},
Sean McGee$^{8}$\orcidlink{0000-0003-3255-3139},
Adam Muzzin$^{9}$\orcidlink{0000-0002-9330-9108},
Julie Nantais$^{10}$\orcidlink{0000-0002-7356-0629},
\newauthor 
Irene Pintos-Castro$^{11}$\orcidlink{0000-0002-9133-4457},
Andrew M. M. Reeves$^{2,3}$\orcidlink{0000-0003-2618-6408},
Gregory H. Rudnick$^{12}$\orcidlink{0000-0001-5851-1856},
\newauthor 
Florian Sarron$^{13}$\orcidlink{0000-0001-8376-0360},
Remco F. J. van der Burg$^{14}$\orcidlink{0000-0003-1535-2327},
Benedetta Vulcani$^{15}$\orcidlink{0000-0003-0980-1499},
\newauthor
Gillian Wilson$^{16}$\orcidlink{0000-0002-6572-7089},
Dennis Zaritsky$^{17}$\orcidlink{0000-0002-5177-727x} \\ \\
\newline \noindent {\normalsize \it Affiliations are listed at the end of the paper}
}

\begin{document}

\pagerange{\pageref{firstpage}--\pageref{lastpage}} 
\pubyear{2022}

\maketitle

\label{firstpage}
\begin{abstract}
We model satellite quenching at $z \sim 1$ by combining $14$ massive ($10^{13.8} < \mhalo/\msun < 10^{15}$) clusters at $0.8 < z < 1.3$ from the GOGREEN and GCLASS surveys with accretion histories of $56$ redshift-matched analogs from the IllustrisTNG simulation.
Our fiducial model, which is parameterized by the satellite quenching timescale ($\tau_{\rm quench}$), accounts for quenching in our simulated satellite population both at the time of infall by using the observed coeval field quenched fraction and after infall by tuning $\tau_{\rm quench}$ to reproduce the observed satellite quenched fraction versus stellar mass trend. 
This model successfully reproduces the observed satellite quenched fraction as a function of stellar mass (by construction), projected cluster-centric radius, and redshift and is consistent with the observed field and cluster stellar mass functions at $z \sim 1$. 
We find that the satellite quenching timescale is mass dependent, in conflict with some previous studies at low and intermediate redshift. 
Over the stellar mass range probed ($\mstar> 10^{10}~\msun$), we find that the satellite quenching timescale decreases with increasing satellite stellar mass from $\sim1.6~{\rm Gyr}$ at $10^{10}~\msun$ to $\sim 0.6 - 1~{\rm Gyr}$ at $10^{11}~\msun$ and is roughly consistent with the total cold gas (H{\scriptsize I}+H$_{2}$) depletion timescales at intermediate $z$, suggesting that starvation may be the dominant driver of environmental quenching at $z < 2$. 
Finally, while environmental mechanisms are relatively efficient at quenching massive satellites, we find that the majority ($\sim65-80\%$) of ultra-massive satellites ($\mstar > 10^{11}~\msun$) are quenched prior to infall.
\end{abstract}


\section{Introduction}
\label{sec:intro}

Observations of galaxies in the local Universe have long shown that various galaxy properties are strongly correlated with the local environment (i.e.~the local galaxy density). For example, satellite galaxies that reside in high-density groups and clusters are more likely to have older stellar populations, exhibit elliptical or spheroidal morphologies, and have depressed rates of star formation relative to their counterparts that reside (primarily as central galaxies) in the lower-density field \citep{Oemler74, Dressler80, Balogh97, Gomez03, Blanton05, Cooper10a}. More recent studies suggest that these environmental trends extend out to $z\sim3$, with passive galaxies already favoring higher-density regions at earlier cosmic times \citep{Cooper06, Cooper07, Cooper10b, Muzzin12, Darvish16, LeeBrown17, Lemaux19, McConachie21}. This distinction between central galaxies that reside in the low-density field and satellite galaxies that reside in high-density groups and clusters may be partially due to the latter population being unable to accrete cold gas after crossing into the virialized region of a group or cluster through a process known as `starvation' or `strangulation' \citep{Larson80, Kawata08}. However, this is far from the only proposed environmentally-driven mechanism for suppressing (or ``quenching'') star formation; other competing mechanisms include ram-pressure stripping \citep{GG72, Abadi99}, tidal stripping \citep{Merritt83, Moore99, Gnedin03}, harassment \citep{Farouki81, Moore96, Moore98}, and feedback-related processes such as overconsumption \citep{McGee14, Balogh16}. Despite the vast number of proposed environmental quenching scenarios, the exact physical mechanism(s) responsible for the aforementioned trends observed in groups and clusters and how they evolve throughout cosmic time remain poorly understood.

A common goal of many studies of environmental (or satellite) quenching is to determine the efficiency with which the local environment suppresses star formation -- i.e.~the timescale upon which satellite quenching operates. 
For that reason, a frequently employed method for understanding quenching efficiency, and potentially isolating the dominant physical mechanism(s) responsible for quenching star formation in dense environments, involves combining observations of groups and clusters with simple quenching models applied to $N$-body simulations to infer the satellite quenching timescale ($\tau_{\rm quench}$), which is typically defined as the time required for a galaxy to transition from star forming to quiescent after becoming a satellite (i.e.~after infall onto its host system).  
A general assumption of this technique is that galaxy quenching can largely be divided into two regimes: [\emph{i}] internal quenching that acts in all environments (or at least within the field population) with increasing efficiency at higher stellar masses and [\emph{ii}] environmental quenching that operates in massive halos or high-density environments (i.e.~groups and clusters) with efficiency that likely depends on local environmental density as well as the mass of the satellite and that of the host halo -- a scenario that is supported by observations at low and intermediate redshift \citep[e.g.][]{Baldry06, Peng10, Woo13, Reeves21}. 
To a large extent, the application of this methodology has primarily been dominated by studies of satellite quenching in the local Universe. In fact, numerous analyses of low-redshift groups and clusters, spanning a broad range in host halo mass, have utilized high-resolution, cosmological 
simulations to infer the typical satellite quenching timescale down to the ultra-faint dwarf regime \citep{Delucia12, Wetzel13, Hirschmann14, Wheeler14, Fham15, Davies16, Pallero19, RodriguezWimberly19}.

Herein, we aim to extend the aforementioned studies of the satellite quenching timescale at low redshift to $z\sim1$ by performing a similar analysis utilizing observations of  satellite galaxies residing in clusters at $0.8 < z < 1.4$. 
In \S\ref{sec:GOGREEN}, we describe our observed galaxy cluster data set, including a discussion of cluster membership criteria and completeness corrections. In \S\ref{sec:TNG}, we detail the high-resolution, cosmological simulation data utilized in our analysis and explain how we construct our simulated sample of cluster galaxies. We describe our satellite quenching model and present the results from implementing said model in \S\ref{sec:Model} and \S\ref{sec:Results}, respectively. Finally, in \S\ref{sec:Discussion}, we discuss variations of our model and how our results relate to similar analyses as a function of cosmic time, before summarizing our results in \S\ref{sec:Conclusion}.
When necessary, we adopt a flat $\Lambda$CDM cosmology with $H_{0} = 70~{\rm km}~{\rm s}^{-1}~{\rm Mpc}^{-1}$ and $\Omega_{m}$ = 0.3. All magnitudes are on the AB system \citep{OkeGunn83}.


\section{Observed Cluster Sample}
\label{sec:GOGREEN}

\subsection{GOGREEN and GCLASS Cluster Sample}
\label{subsec:2.1}

Our cluster sample is drawn from the Gemini CLuster Astrophysics Spectroscopic Survey (GCLASS) and the Gemini Observations of Galaxies in Rich Early ENvironments (GOGREEN) survey \citep{Muzzin12, Balogh17, Balogh21}.\footnote{http://gogreensurvey.ca/data-releases/data-packages/gogreen-and-gclass-first-data-release/} These surveys combine deep, multi-wavelength photometry with extensive Gemini/GMOS \citep{Hook04} spectroscopy of galaxies in 26 overdense systems over a redshift range of $0.867 \lt z \lt 1.461$, with the primary objective of studying galaxy evolution in high-density environments. The sample utilized in our analysis consists of $14$ clusters with halo masses in the range from $10^{13.8-15}~\msun$ and spectroscopic redshifts of $0.867 \lt z  \lt 1.368$. Eleven of these clusters were selected from the Spitzer Adaptation of the Red-sequence Cluster Survey \citep[SpARCS,][]{Wilson09, Muzzin09, Demarco10}, where they were detected in shallow $z'$ and IRAC $3.6\mu$m images due to their overdensity of red-sequence galaxies \citep{GladdersYee00}. The remaining three clusters were drawn from the South Pole Telescope (SPT) survey  \citep{Brodwin10, Foley11, Stalder13} and were initially detected via their Sunyaev-Zeldovich \citep{SunyaevZeldovich70} signature and later spectroscopically confirmed. In Table~\ref{table:1}, we provide properties of our cluster sample including halo mass ($M_{200}$) and radial scale ($R_{200}$) -- which are both obtained using the MAMPOSSt method \citep{MBB13} as outlined in \citet{Biviano21} -- along with redshift and the number of spectroscopic cluster members with $\mstar >10^{10}~\msun$.

\begin{table}
\centering
\begin{tabular}{ccccc}
    \hline
    \hline  
    
    \multirow{2}{*}{Name}  &  $M_{200}$  & $R_{200}$ & \multirow{2}{*}{$z$} & \multirow{2}{*}{$N_{\rm members}$} \\
     & [$10^{14}~{\rm M}_{\odot}$] & $[\rm cMpc]$ &  &  \\

    \hline

    SpARCS0034 &   0.6 &   1.08   &  0.867 &  23\\
    SpARCS0035 &    3.8 &   2.17   &  1.335 &  18\\
    SpARCS0036 &    3.6 &  2.09   &  0.869 &  45\\
    SpARCS0215 &    2.4 &   1.70   &  1.004 &  34\\
    SpARCS0335 &    1.8 &   1.59   &  1.368 &  7\\
    SpARCS1047 &    2.5 &   1.78   &  0.956 &  26\\
    SpARCS1051 &    2.2 &   1.80   &  1.035 &  26\\
    SpARCS1613 &    11.1 &   2.97   &  0.871 &  68\\
    SpARCS1616 &    3.3 &   1.98   &  1.156 &  39\\
    SpARCS1634 &    2.7 &   1.85   &  1.177 &  34\\
    SpARCS1638 &    1.7 &  1.56   &  1.196 &  20\\
    SPT0205    &    3.1 &   1.77   &  1.323 &  19\\
    SPT0546    &    5.8 &   2.42   &  1.067&  27\\
    SPT2106    &    7.3 &   2.62   &  1.131 &  30\\

    \hline
    \hline
    \label{table:obs_galaxies}
    \label{table:1}
\end{tabular}
\caption{Properties of our GOGREEN cluster sample, including $\mtwo$, $R_{200}$, cluster redshift, and the number of spectroscopic members (with $\mstar >10^{10}~\msun$). The values in the $R_{200}$ and $\mtwo$ columns were obtained using the MAMPOSSt method \citep{MBB13} as outlined in \citet{Biviano21}. Details regarding the cluster membership criteria are discussed in Sec.~\ref{subsec:2.2}.}
\end{table}

We also utilize data from the deep, multi-wavelength imaging of each GOGREEN system \citep{vdB13, vdB20}. From the photometric catalogs, we employ photometric redshift and stellar mass measurements as well as rest-frame $U-V$ and $V-J$ colors, which are used to determine cluster membership and classify galaxies as either star forming or quenched (see \S\ref{subsec:2.2}). As described in \cite{vdB20}, the photometric redshifts were estimated using the \texttt{EAZY} code (Version May 2015, \citealt{Brammer08}) by fitting the multi-wavelength photometry to spectral energy distribution templates from the PEGASE model library \citep{FiocRocca97} along with a red galaxy template from \cite{Maraston05}. Furthermore, the stellar masses were estimated by fitting the photometry to stellar population synthesis models \citep{BruzualCharlot03} using the \texttt{FAST} code \citep{Kriek09}, assuming solar metallicity, a \cite{Chabrier03} initial mass function, and the dust law from \cite{Calzetti00}.

\subsection{GOGREEN Cluster Membership and Classification}
\label{subsec:2.2}

We determine cluster membership for our observational sample by first measuring the comoving projected radial cluster-centric distance, $R_{\rm proj}$, for all objects -- excluding the centrals -- in the field of the $14$ clusters that comprise our sample. We then exclude all objects that are not within $R_{200}$ of the cluster, which is defined as the comoving radius of a sphere centered at the position of the central within which the mean density is 200 times the critical density of the Universe. We further restrict our satellite sample to only include objects with $\mstar > 10^{10}~\msun$, which is slightly above the $\sim 80\%$ stellar mass completeness limit for the sample \citep{vdB20}. 
From here, we apply the following cluster membership selection criteria to the subsample of objects with high-quality spectroscopic redshifts. Namely, we only include objects with secure spectroscopic redshifts (Redshift\_Quality = 3,4) and $\lvert z_{\rm{spec}} - z_{\rm{cluster}} \rvert  \leq 0.02(1+z_{\rm{spec}})$.\footnote{Please refer to \citet{Balogh21} for a description of the redshift quality flags and the assignment process.} Likewise, for the subsample of objects without high-quality spectroscopic redshifts, we identify membership based on objects with STAR $\neq$ 1 and $\lvert z_{\rm{phot}} - z_{\rm{cluster}} \rvert  \leq 0.08(1+z_{\rm{phot}}$), where the STAR flag is the GOGREEN star/galaxy classification based on color selection, as described in \cite{vdB20}. The choice to only include galaxies with  $\lvert z_{\rm{phot}} - z_{\rm{cluster}} \rvert  \leq 0.08(1+z_{\rm{phot}})$ was informed by our knowledge that the photometric-redshift uncertainty for galaxies more massive than $10^{10}~\msun$ is $0.048(1+z)$. Nevertheless, we find that if we subsequently characterize and account for interlopers and incompleteness, as described in \S\ref{subsec:2.3}, the results of our analysis do not depend on the 
 
$\Delta z$ threshold adopted as part of 
this particular membership criterion. Altogether, these membership selection criteria yield a total of 1072 cluster members (416 spectroscopic/656 photometric). Lastly, we classify the quiescent members of our cluster population using the following rest-frame $UVJ$ color-color cuts defined by \citet[][see also \citealt{Williams09}]{Whitaker11}:
\begin{equation}
\begin{split}
(\mathrm{U-V}) > 1.3~\cap~(\mathrm{V-J}) \lt 1.6~\cap~  \\ 
(\mathrm{U-V}) > 0.88 \times (\mathrm{V-J}) +  0.59  \; .
\end{split}
\label{uvj_eqn}
\end{equation}

\subsection{Completeness Correction}
\label{subsec:2.3}

In order to obtain an accurate measurement of the satellite quenched fraction, we must account for incompleteness and interlopers that inevitably contaminate our photometric sample. This is accomplished following the methodology utilized in \cite{vdB13, vdB20} that accounts for completeness in the cluster sample by computing a membership correction factor using the sample of galaxies with both multi-band photometry and $z_{\rm spec}$ measurements and then applying that factor to the photometric sample. The membership correction factor (Eqn.~\ref{c_factor}) is defined as the sum of the number of galaxies that are either secure cluster members or false negatives divided by the sum of secure cluster members and false positives, 
\begin{equation}
\label{c_factor}
\textit{C}_{\rm{factor}} = \frac{\textit{N}(\rm{secure~cluster}) + \textit{N}(\rm{false~negative})  }{\textit{N}(\rm{secure~cluster})+ \textit{N}(\rm{false~positive})} \; .
\end{equation}
Here, secure cluster members are defined as objects identified as cluster members based on their spectroscopic redshift \emph{and} with photometric redshifts consistent with membership, whereas false negatives are objects that are spectroscopically-confirmed cluster members with a photo-$z$ that is inconsistent with cluster membership. Lastly, false positives are defined as objects that are not cluster members based on their spectroscopic redshift but have a photo-$z$ consistent with the redshift of the cluster. 
Following the methodology of \citet{vdB20}, we compute the correction factor separately for star-forming and quiescent galaxies in order to account for the presumed color dependence of field contamination. Furthermore, for both populations we compute the correction factor in bins of stellar mass ranging from $10^{10.0-11.4}~\msun$ and $R_{\rm{proj}}/\rtwo$ from $0-1$. 
As a function of galaxy color, we find a very modest variation in the completeness correction, with the correction factor as applied to the star-forming and quiescent populations differing by $\lesssim2\%$.
Finally, we apply the appropriate correction factor as a weight to each cluster member, which we find yields a modest change in the measured quenched fractions (at the level of $\sim1-2.5\%$), such that the final results of our analysis and the conclusions therein drawn remain unchanged irrespective of the application of this completeness correction.

\section{Simulated Cluster Sample}
\label{sec:TNG}

\subsection{IllustrisTNG Cluster Sample}
\label{subsec:3.1}

We utilize the TNG300-1 simulation from the IllustrisTNG project\footnote{https://www.tng-project.org} \citep[TNG,][]{Nelson18, Niaman18, Springel18, Pillepich18, Marinacci18} to establish a simulated cluster population that is matched on redshift to our observed cluster sample. TNG300-1 is a large volume ($\sim300~{\rm cMpc}^{3}$), high-resolution ($2 \times 2500^{2}$ resolution elements), cosmological, gravo-magnetohydrodynamical simulation that utilizes the moving mesh \texttt{AREPO} code and solves for the coupled evolution of dark matter, cosmic gas, luminous stars, and supermassive black holes from a starting redshift of $z=127$ to the present day, $z=0$. TNG300-1 has a dark matter (gas) mass resolution of $m_{\rm DM} = 5.9 \times 10^{7}~\msun$ ($m_{\rm baryon} = 1.1 \times 10^{7}~\msun$), which corresponds to a halo mass (stellar mass) completeness of $\sim10^{10}~\msun$ ($\sim10^{9}~\msun$). As explained in \S3.3 of \cite{Pillepich18}, we augment the stellar masses for TNG300-1 galaxies at $z\sim 1$ by a factor of $1.3\times$ to account for resolution limitations that systematically underestimate stellar masses within the simulations.


\begin{figure}
 \centering
 \hspace*{-0.2in}
 \vspace*{-0.2in}
 \includegraphics[width=3.5in]{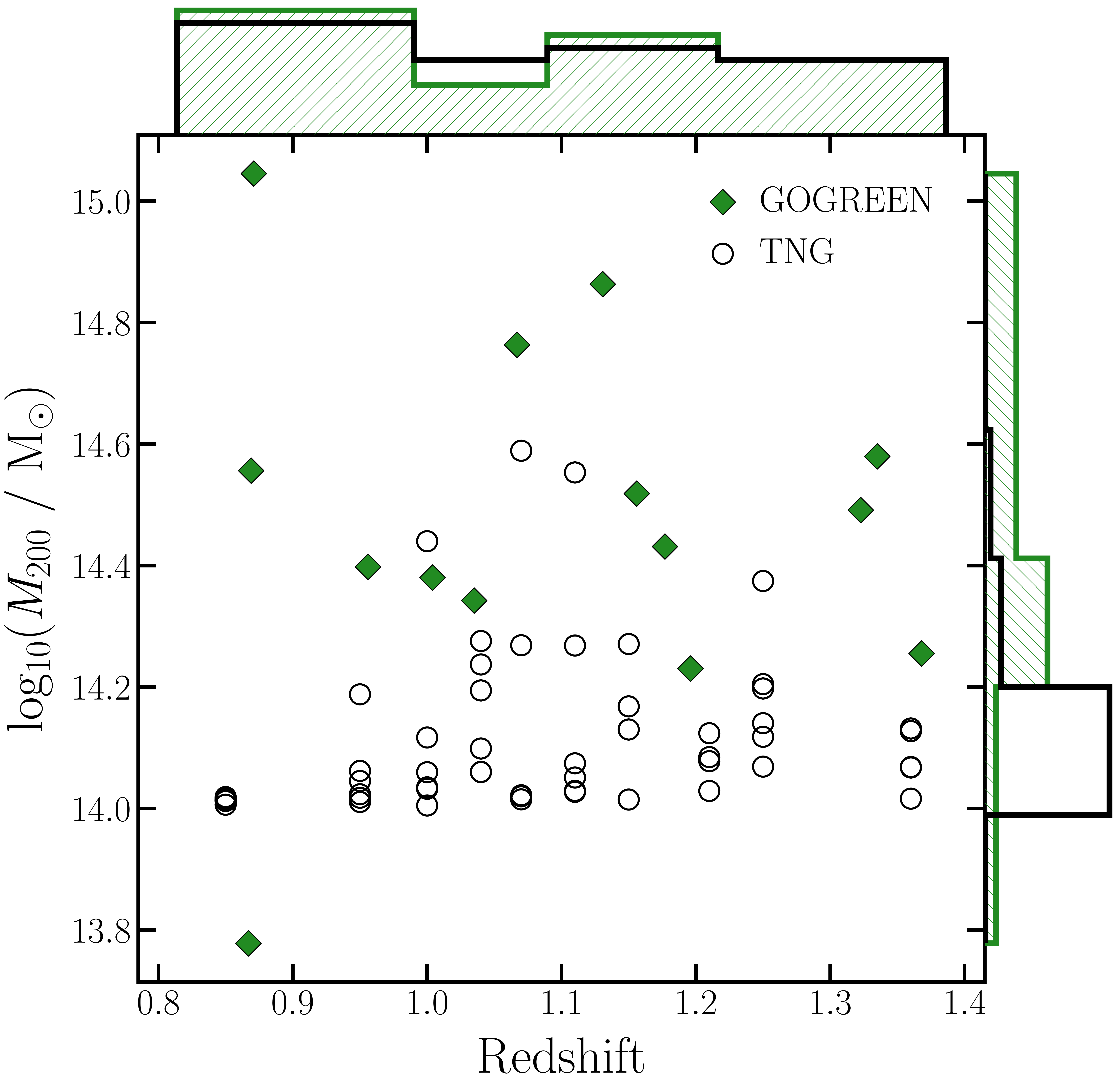}
 \caption{$\mtwo$ versus $z$ for the observed and simulated cluster samples. The open circles (filled diamonds) represent the TNG (GOGREEN) clusters. While matched on redshift, the simulated sample is biased towards less-massive systems relative to the observed sample, with the majority of the TNG clusters having halo masses less than $10^{14.3}~\msun$. As discussed in \S\ref{subsec:3.1}, this bias towards low-mass hosts does not significantly impact our results, with a sample matched on $\mtwo$ yielding qualitatively similar results.}
 \label{fig:m200_vs_z}
\end{figure}


\begin{figure}
 \centering
 \hspace*{-0.25in}
 \includegraphics[width=3.5in]{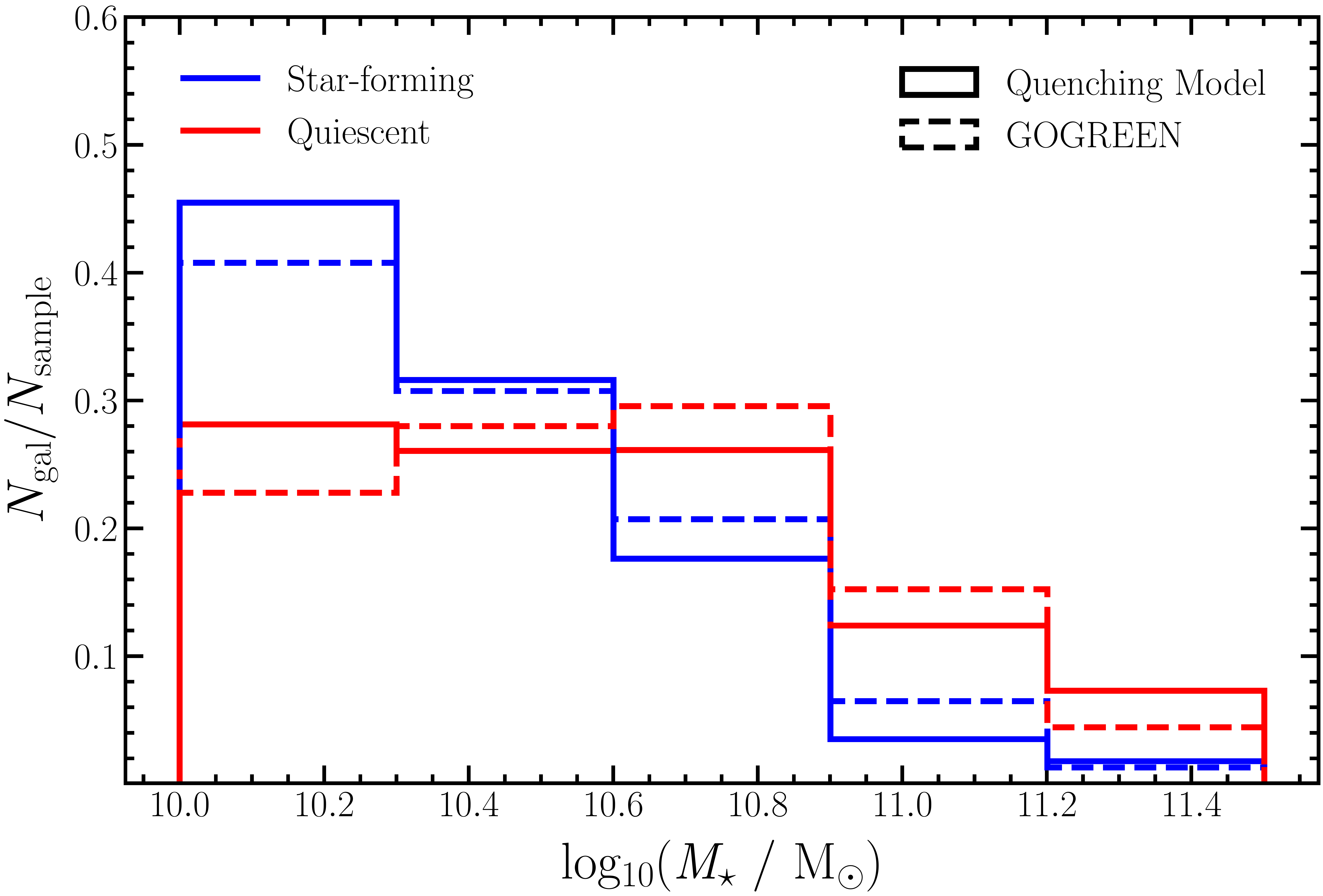}
 \caption{Comparison of the normalized stellar mass distributions for GOGREEN cluster members to that of our simulated satellite population. The blue and red solid (dashed) lines illustrate the simulated (observed) stellar mass distribution for star-forming and quenched galaxies, respectively. Note that the simulated cluster members are classified according to our quenching model that is designed to reproduce the observed $f^{sat}_{\rm{q}}(\mstar)$ results (see \S\ref{sec:Model}). While the TNG sample slightly underpredicts the total number of satellites due its bias towards lower host halo masses, the relative distribution of satellite masses is in excellent agreement.}
 \label{fig:mstar_dist}
\end{figure}


Our simulated cluster sample is drawn from the group catalogs and sublink merger trees associated with the TNG300-1 simulation. As a whole, TNG300-1 contains a total of 100 snapshots ranging from $z=20.05$ to $z=0$; however, our cluster sample is constructed using only 10 snapshots ranging from $z=1.36$ to $z=0.85$, so as to match the redshift distribution of the GOGREEN cluster sample. Each of these snapshots contains a unique group catalog that includes both friends-of-friends \citep[FoF;][]{Davis85} and \textsc{Subfind} objects \citep{Springel01, Dolag09}. The FoF catalog contains the GroupFirstSub column that holds the indices into the \textsc{Subfind} catalog for the first/primary/most massive subhalo group within each FoF group, and we define these subhalos to be our centrals. 
With the total central population defined, we use the TNG300-1 Sublink merger trees to track the \textsc{Subfind} IDs of the sample from $z=0.85$ to $z=1.36$, which allows unique centrals to be identified across the $10$ snapshots. Moreover, we combine this information with the redshift distribution of our observed cluster population to construct a sample of simulated clusters that is matched on redshift to the GOGREEN cluster sample. Given the relatively large volume of the TNG300-1 simulation box, we are able to select a total of $56$ unique comparison cluster halos from snapshots that range from $z=1.36$ to $z=0.85$ with a median redshift of $z=1.1$. The median redshift difference between a GOGREEN cluster and its simulated analog is $|\Delta z| \sim 0.03$.
As illustrated in Fig.~\ref{fig:m200_vs_z}, our simulated host sample has a median halo mass of $\mhalo = 10^{14.12}~\msun$ and is, on average, less massive than the GOGREEN cluster sample, which has a median host mass of $10^{14.5}~\msun$. A consequence of this is that the number of simulated cluster members in our sample is generally less than their observed counterparts by a factor of $\sim 3$. With this in mind, we repeat our analysis using a more restricted sample of $12$ clusters constructed to better match the observed GOGREEN cluster sample with respect to redshift, halo mass, and $R_{200}$. Utilizing this more-precisely matched sample, we find that our results are qualitatively similar to those based on the the larger and less-precisely matched sample. The robustness of our results is, in part, due to the fact that at fixed stellar mass the infall time distribution for satellites in the low-mass and high-mass clusters, a key ingredient in our modeling (see \S\ref{sec:Model}), is weakly dependent on host mass with differences in average infall times on the order of $\sim0.02-0.03~{\rm Gyr}$. All things considered, we choose the larger host sample, matched solely on redshift, as our simulated cluster population in part due to its ability to better sample the distribution of infall times (and formation histories).

\subsection{TNG Cluster Membership}
\label{subsec:3.2}

For each of the simulated clusters, our sample of cluster members is drawn from the TNG300-1 group catalogs and sublink merger trees. In particular, we define potential cluster members as any object in the Subfind catalog that is not defined as the host within each FoF group. From here, we establish cluster membership for our simulated cluster sample using a procedure similar to that outlined in \S\ref{subsec:2.2}. Specifically, simulated cluster members are galaxies that satisfy the condition $d_{\rm{host}}(z_{\rm{obs}}) \lt R_{200}$, where $d_{\rm{host}}(z_{\rm{obs}})$ is the three-dimensional comoving radial cluster-centric distance at the redshift of observation.
We note that this satellite selection criterion is distinct from how observational samples are selected, where projected separations are typically utilized given that three-dimensional separations are largely unattainable. For this reason, we repeat our analysis using a cluster member sample composed of galaxies that lie within a cylinder of radius $\rtwo$ projected on an imaginary sky plane perpendicular to the z-direction of the simulation box, which we define as the
line-of-sight direction. In general, we find that selecting satellites according to projected cluster-centric distance yields consistent, though slightly shorter quenching timescales, with the difference (relative to selecting in 3-D) being most pronounced at low satellite masses ($\Delta\tau_{\rm quench}\lesssim -0.1$~Gyr). We find that this remains true even if the satellite selection criterion is expanded to include a line-of-sight velocity threshold analogous to the $\Delta z$ threshold used for the observed satellite sample. The weak bias towards shorter quenching timescales, when working in projected space, is primarily driven by the inclusion of star-forming interlopers from the field population \citep{Donnari21}.

In addition to the separation criterion, we also restrict our simulated satellite sample to only include galaxies with resolution-corrected stellar mass of $\mstar > 10^{10}~\msun$, where the stellar masses are given by the total mass of all star particles associated with each galaxy (i.e. IllustrisTNG SubhaloMassType masses with Type=4). Our adopted stellar mass limit, selected to mirror that of the GOGREEN sample, is well above the stellar mass completeness limit for TNG300-1 of approximately $\mstar \sim 10^{9}~\msun$, which corresponds to $\sim100$ star particles. Overall, these constraints yield a total of 1220 cluster members across the 56 simulated clusters.
As illustrated in Figure~\ref{fig:mstar_dist}, the TNG-based stellar masses reproduce the relative distribution of satellite stellar masses from the GOGREEN sample. 
The stellar masses for the simulated satellite sample are taken at $z_{\rm obs}$, such that we do not explicitly model the stellar mass growth of satellites prior to or following infall. 
The difference in mass due to subsequent star formation (or lack thereof) in comparison to the star formation histories defined by the TNG hydro-dynamical modeling is modest (typically $\Delta \mstar \lesssim 0.3~{\rm dex}$).
In lieu of using the stellar masses provided by TNG300-1, we discuss the implications of defining the stellar masses of our cluster satellites using the stellar mass-halo mass (SMHM) relation from \citet{Behroozi13} in \S\ref{subsec:mstar_impact}. 
Finally, after establishing the simulated galaxy sample we proceed to use the TNG300-1 sublink merger trees to track relevant properties (e.g. position, mass, $R_{200}$, etc.) of the clusters and their members along the main progenitor branch from $z=20.05$ to $z_{\rm obs}$.


\begin{figure}
 \centering
 \hspace*{-0.25in}
 \includegraphics[width=0.99\linewidth]{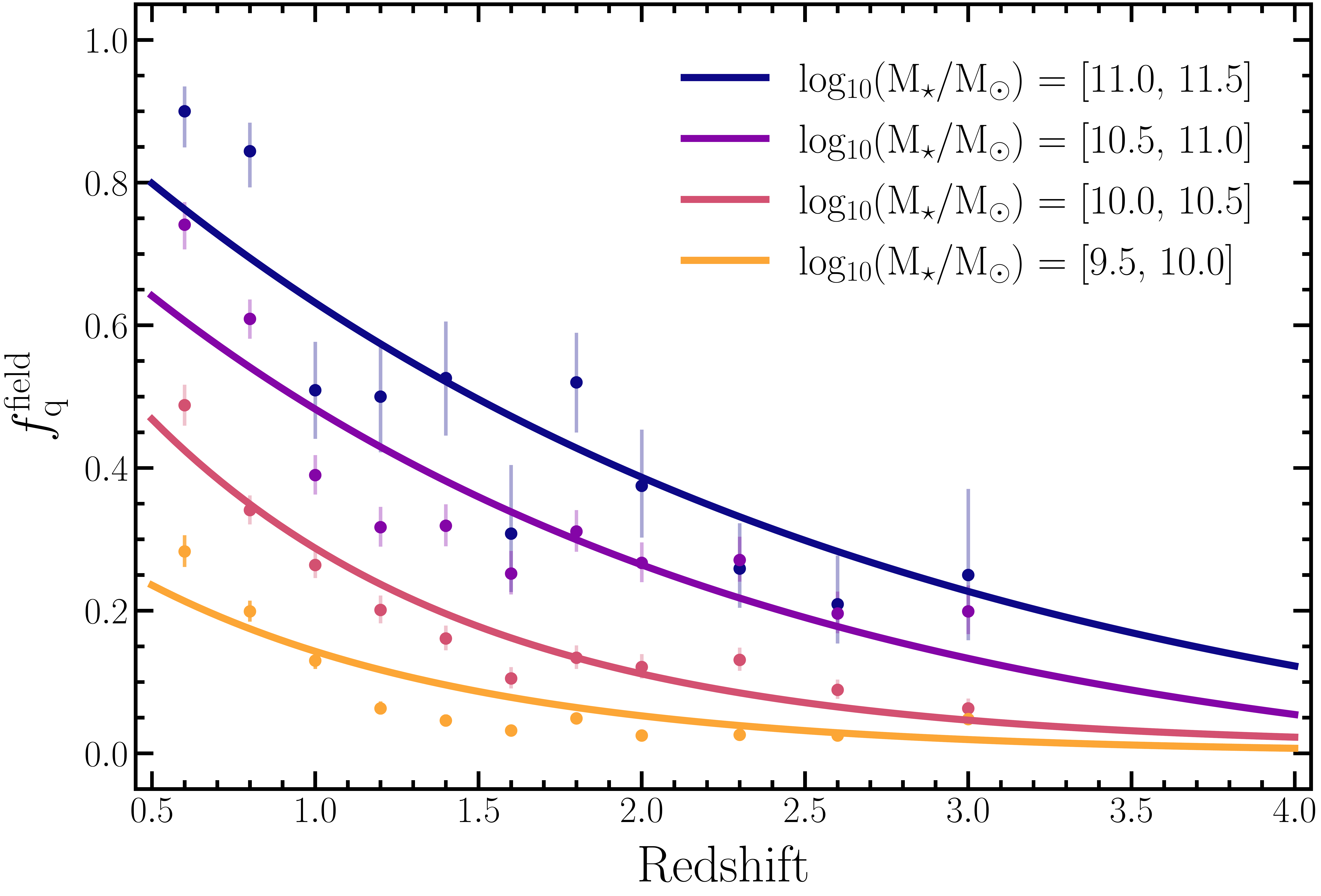}
 \caption{Field quenched fraction versus redshift in bins of stellar mass ranging from $10^{9.5}~\msun \lt \mstar \lt 10^{11.5}~\msun$ as inferred from CANDELS observations. The colored circles represent the observed field quenched results in their respective stellar-mass bins, whereas the curves illustrate the corresponding fits to the observed results using an exponentially decaying function. The vertical error bars correspond to the 1-$\sigma$ binomial uncertainties in the quenched fraction.}
 \label{fig:CANDELS_fq_vs_mstar}
\end{figure}

\section{Quenching Model}
\label{sec:Model}

Our quenching model utilizes the TNG simulations to detail the accretion history of the cluster population and complementary ``field'' observations to describe the properties of infalling galaxies. Together, these inputs allow the model to 
probabilistically characterize galaxies that quenched prior to infall onto the cluster using the coeval field quenched fraction. 
At its core, the model has one primary parameter, the satellite quenching timescale ($\tau_{\rm quench}$), which is defined as the time following infall before a star-forming satellite is quenched. 
This model parameter is tuned so as to reproduce the observed dependence of the satellite quenched fraction on stellar mass, $f^{\rm sat}_{{\rm q}}(\mstar)$, thereby yielding $\tau_{\rm{quench}}(\mstar)$.

\subsection{Infall Times of Simulated Cluster Members}
\label{subsec:4.1}
 
Our procedure for classifying the simulated cluster members that quenched prior to infall begins with computing the infall time ($t_{\rm{infall}}$) for each simulated satellite, which we define as the time at which a galaxy first crosses $R_{200}$ of the cluster halo. For our simulated satellite population, less than 20\% are backsplash systems that crossed $R_{200}$ more than once, with $t_{\rm infall}$ defined as the time of the first crossing.
As discussed in \S\ref{subsec:intq}, we also investigate an alternative approach in which we classify simulated cluster members at the redshift of observation (versus at the time of infall) to account for the possibility of internal quenching after infall. 
To measure $t_{\rm infall}$, we use the TNG300-1 sublink merger trees (see \S\ref{subsec:3.2}) to track the separation between our simulated cluster and satellite samples across the $55$ snapshots between $z=20.05$ and $z=0.85$. This corresponds to a median time resolution of approximately $100$ Myr between each snapshot, which is not ideal for precisely measuring $t_{\rm{infall}}$ given that the radial cluster-centric separation can change on the order of a few hundred kpc between each snapshot. Therefore, with the objective of obtaining greater precision on $t_{\rm{infall}}$, we map the spatial position of each galaxy (relative to their host cluster halo) in 10 Myr intervals by spline interpolating the position of each galaxy and corresponding host from $z=20.05$ to the redshift of the given snapshot. We find that the infall times procured using the spline-interpolated positions are typically $\sim 60$~Myr earlier when compared to the infall times obtained using the non-interpolated positions. In the following section, we explain how we use these infall times to probabilistically classify our simulated cluster members as star forming or quiescent.

\begin{figure*}
 \centering
 \hspace*{-0.1in}
 \includegraphics[width=7.0in]{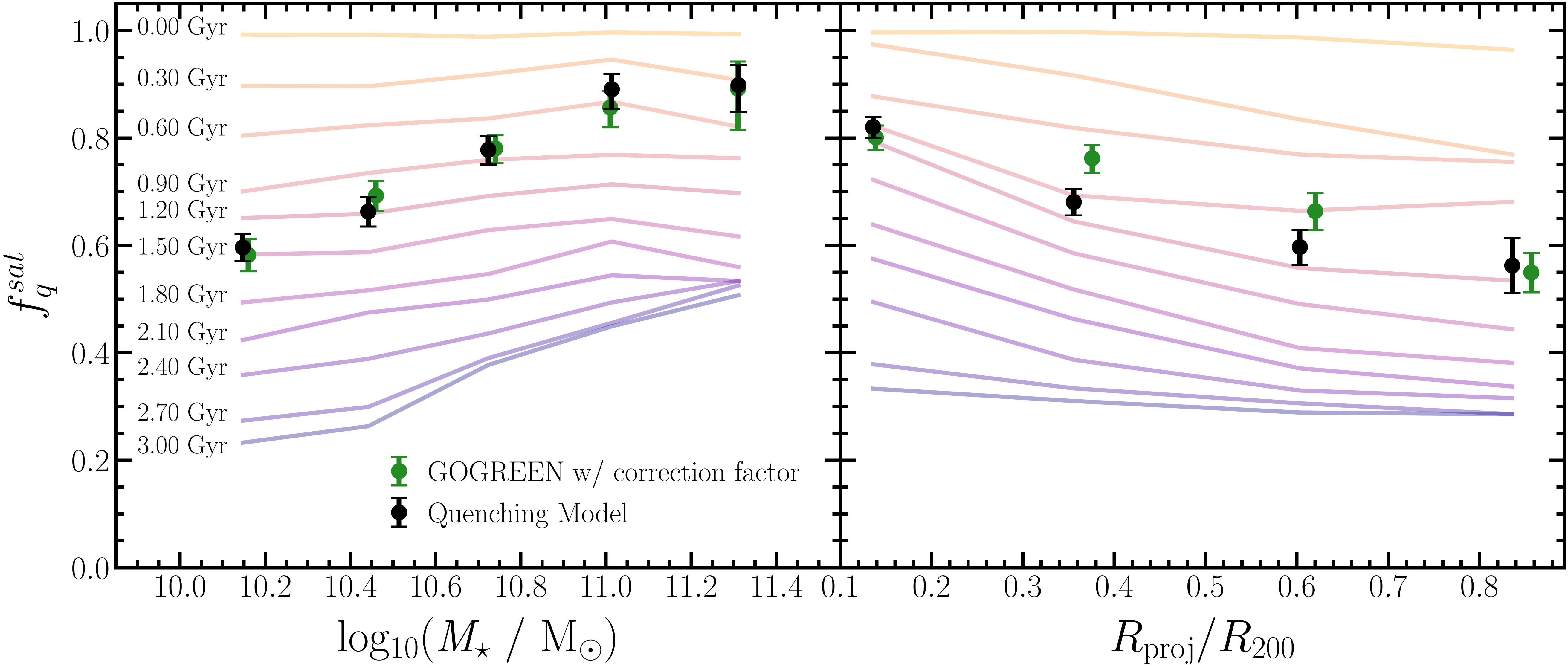}
 \caption{Satellite quenched fraction as a function of satellite stellar mass ({\it left}) and projected cluster-centric distance normalized by $\rtwo$ ({\it right}). The green circles illustrate the GOGREEN quenched fraction results with the membership correction factor applied. The black circles represent the TNG results fit to the GOGREEN quenched fraction results. The colored profiles in the background represent the TNG quenched fraction results using a constant quenching timescale ranging from 0 to 3 Gyr. The constant quenching timescale model fails to reproduce the observed quenched fraction as a function of stellar mass and cluster-centric radius, however, these trends are reproduced by a model assuming a mass-dependent quenching timescale. All error bars correspond to the 1-sigma binomial uncertainties.}
 \label{fig:fq_vs_mstar}
\end{figure*}


\subsection{Classifying Simulated Cluster Members}
\label{subsec:4.2}

Within our satellite quenching model, each infalling system is probabilistically classified as star forming or quenched according to the corresponding field quenched fraction at the time of infall. 
In Figure \ref{fig:CANDELS_fq_vs_mstar}, we show the field quenched fraction as a function of redshift and stellar mass, $f^{\rm field}_{{\rm q}}(z, \mstar)$, computed using derived data products from the v1.1 internal data release of the Cosmic Assembly Near‑infrared Deep Extragalactic Legacy Survey \citep[CANDELS,][]{Grogin11, Koekemoer11, Guo13, Galametz13, Santini15, Stefanon17, Nayyeri17, Barro19}. To obtain the field quenched fraction, we first identified objects in the CANDELS catalogs with reliable photometry (PHOTFLAG==0) and identified the fraction in the quiescent region of the $UVJ$ diagram following \citet{Whitaker11}. Our field sample totals $57,971$ galaxies, with each bin in redshift and mass 
including no fewer than $20$ galaxies. 
In agreement with previous analyses, we find that the field quenched fraction depends strongly on stellar mass and redshift, with more massive galaxies more likely to be quenched and the prevalence of quenched systems decreasing at earlier cosmic time. 
We also find that corresponding measurements of the field quenched fraction, computed using a $K_{s}$-selected catalog drawn from the COSMOS/UltraVISTA field \citep{Muzzin13a, Muzzin13b, Marsan22}, yield results that are generally consistent with those derived from the CANDELS dataset. 

As previously mentioned, we use $f^{\rm field}_{{\rm q}}(z, \mstar)$ to probabilistically classify the simulated cluster members that quenched prior to infall. We accomplish this by first fitting the measurements of the field quenched fraction in mass bins (see Fig.~\ref{fig:CANDELS_fq_vs_mstar}) using an exponentially decaying function to obtain functional forms for the four stellar mass bins between $10^{9.5-11.5}~\msun$. We then use $z_{\rm infall}$ and $\minfall$ values of our simulated satellite population to obtain the expected field quenched fraction at the time of infall. Next, we randomly draw a number from a uniform distribution between zero and one and compare it with the corresponding field quenched fraction. If the randomly drawn number is greater (less) than the observed quenched fraction then we classify the galaxy as star forming (quenched). This step is repeated 50 times in order to generate an ensemble of classified cluster members that capture the slight variations inherent to this probabilistic classification scheme. As such, the quenched fraction results discussed in \S\ref{subsec:5.1} represent the median of the ensemble of classified cluster members.  
\subsection{Determining the Satellite Quenching Timescale}
\label{subsec:4.3}

We characterize environmental quenching by employing a simple quenching model that assumes that star-forming satellites quench after some fixed amount of time ($\tau_{\rm quench}$) following infall onto their host cluster halo. The simplicity of this model is that it contains one primary parameter, $\tau_{\rm quench}(\mstar)$, which we allow to vary  linearly with satellite stellar mass so as to reproduce the $f^{\rm sat}_{{\rm q}}(\mstar)$ measurements for our observed cluster sample. In other words, our model translates the observed $f^{\rm sat}_{{\rm q}}(\mstar)$ into typical quenching timescales by inferring $\tau_{\rm quench}$ in bins of stellar mass so as to minimize the difference between the model and the observations ($\lvert f_{\rm{q, obs}}(\mstar) - f_{\rm{q, model}}(\mstar) \rvert$). 
For an infinitely-long quenching timescale (i.e.~no environmental quenching), the minimum satellite quenched fraction is defined by the portion of satellites quenched prior to infall.
In \S\ref{sec:Results}, we present the results of our environmental quenching model and discuss the implications of the inferred quenching timescales.

\section{Results}
\label{sec:Results}

\subsection{Quenched Fraction Results}
\label{subsec:5.1}

In Figure \ref{fig:fq_vs_mstar}, we compare the GOGREEN observed satellite quenched fraction as a function of stellar mass and projected cluster-centric distance with the corresponding quenched fraction results from our environmental quenching model. The green circles represent the observed results with the membership correction factor applied. As noted in \S\ref{subsec:2.3}, the membership correction factor has a relatively small impact on the observed quenched fraction results. 
We find a strong dependence of the quenched fraction on both $\mstar$ and $R_{\rm proj}/R_{200}$, such that more massive and more centrally-located satellites are more likely to be quenched.  
These observed trends are in good agreement with similar results at low and intermediate redshift \citep[e.g.][]{Balogh98, CZ05, Patel09, Vulcani15, Cooke16, LeeBrown17, Baxter21}.
The faded colored lines in Fig.~\ref{fig:fq_vs_mstar} show the simulated quenched fraction results when assuming a constant quenching timescale (independent of satellite stellar mass), ranging from $\tau_{\rm quench} = 0-3~{\rm Gyr}$. As illustrated, a fixed quenching timescale fails to reproduce the observed satellite quenched fraction versus stellar mass trend.
In contrast, the results of our fiducial quenching model, which assumes a mass-dependent satellite quenching timescale, are illustrated by the black circles in Fig.~\ref{fig:fq_vs_mstar}. While our model yields the observed $f_{{\rm q, sat}}(\mstar)$, by design, it also successfully reproduces the observed dependence of quenched fraction on projected cluster-centric distance within the GOGREEN cluster sample.

\begin{figure}
 \centering
 \hspace*{-0.25in}
 \includegraphics[width=1.0\linewidth]{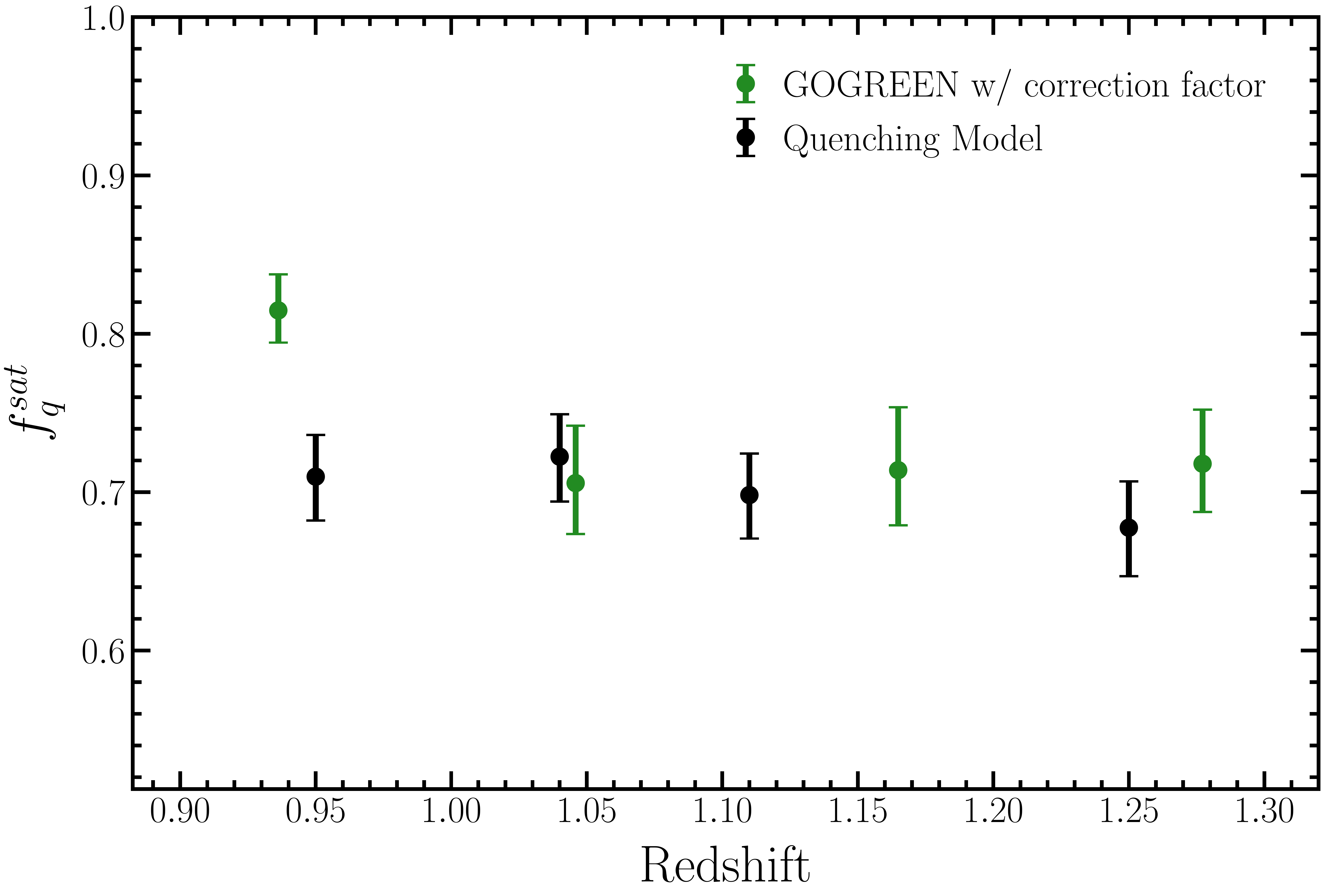}
 \caption{Satellite quenched fraction versus redshift. The green circles represent the observed results with the membership correction applied. The black circles shows the corresponding measurements for our fiducial model based on tuning $\tau_{\rm{quench}}(\mstar)$ to reproduce the observed satellite quenched fraction as a function of stellar mass. For both the observed and simulated samples, the uncertainties correspond to 1-$\sigma$ binomial errors. Our fiducial quenching model is able to successfully reproduce the observed GOGREEN satellite quenched fraction as a function of stellar mass, projected cluster-centric radius, \emph{and} redshift.}
 \label{fig:fq_vs_z}
\end{figure}

In Figure~\ref{fig:fq_vs_z}, we compare the observed satellite quenched fraction as a function of redshift to the results from our fiducial quenching model. Over the limited redshift range probed by the GOGREEN survey, the measured satellite quenched fraction is relatively constant (see also \citealt{Nantais17}), with excellent agreement between the results for the observed and simulated cluster samples. 
Overall, our fiducial quenching model is extremely successful, reproducing the observed quenched fraction as a function of stellar mass (by construction), projected cluster-centric distance, \emph{and} redshift.

\subsection{Inferred Quenching Timescales}
\label{subsec:tau}

\begin{figure*}
 \centering
 \hspace*{-0.1in}
 \includegraphics[width=0.8\linewidth]{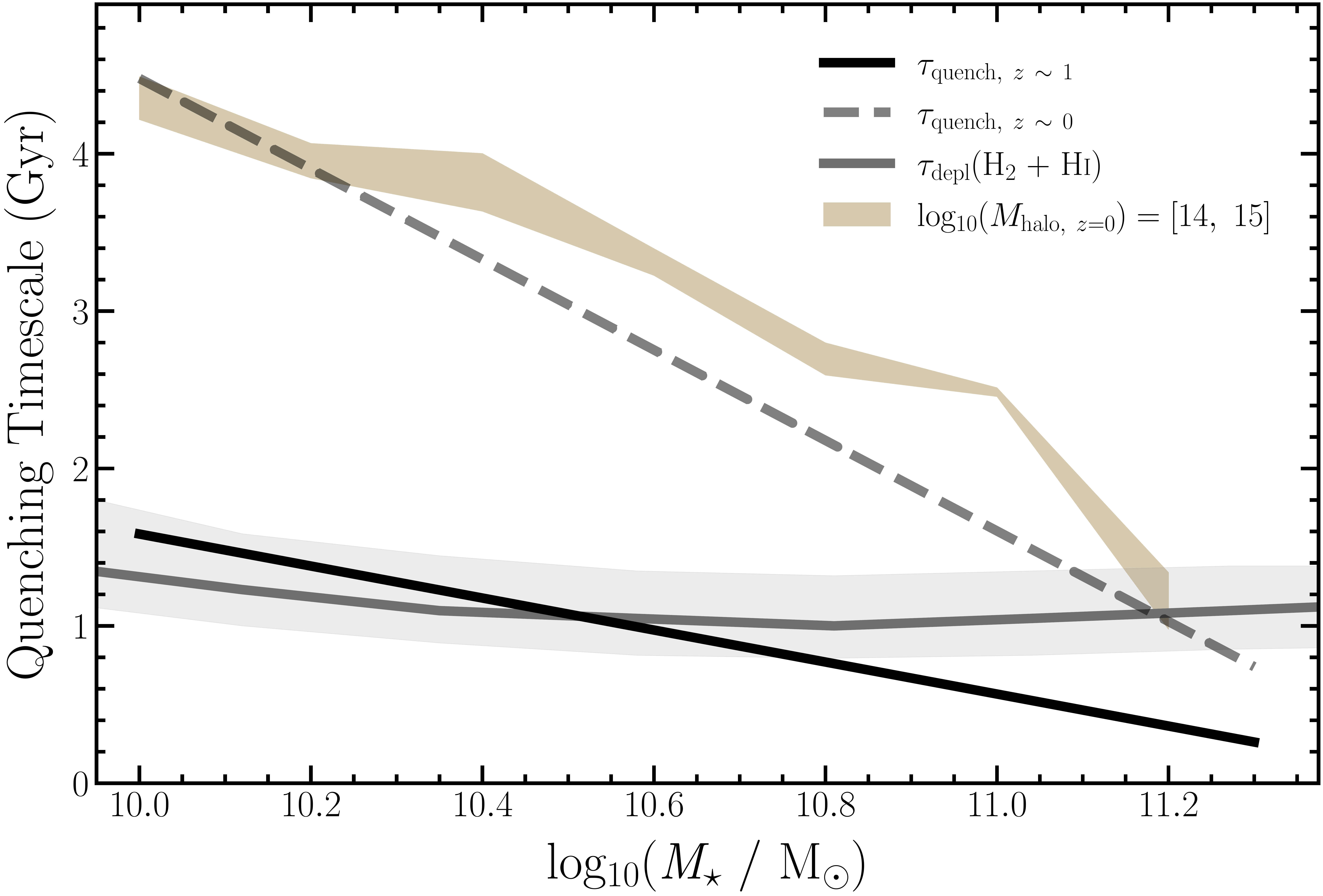}
 \caption{Satellite quenching timescale versus satellite stellar mass. The solid grey line illustrates the empirically-derived cold gas (H{\scriptsize I} + H$_2$) depletion timescale from \citet{Popping15} at $z \sim 1.5$, with the corresponding grey shaded region spanning the variation in the depletion timescale over the redshift range $1 < z < 2$. The solid black line represents the results from our fiducial model as applied to the GOGREEN cluster sample at $z \sim 1$ ($\mhalo \sim 10^{14.5}$). The dashed grey line represents the estimated quenching timescale at $z \sim 0$ obtained by scaling the results from our fiducial model at $z \sim 1$ by $(1+z)^{-3/2}$. In our fiducial model, we find a mass-dependent quenching timescale, favoring more rapid suppression of star formation for more massive satellites. For comparison, the tan colored band shows the quenching timescale constraint from \citet{Wetzel13} for satellites in clusters ($\mhalo \sim 10^{14-15}~\msun$) at $z\sim 0$. For massive hosts, the evolution in the quenching timescale roughly follows the evolution in the dynamical time ($\times(1+z)^{-3/2}$), as shown by the dashed grey line.}
  \label{fig:tq_vs_mstar}
\end{figure*}

In Figure~\ref{fig:tq_vs_mstar}, we present the $\tau_{\rm quench}(\mstar)$ results that we infer from our fiducial environmental quenching model. Within the framework of our modeling approach, we find that a mass-dependent quenching timescale in which higher-mass galaxies quench more rapidly following infall onto their host halo is necessary to reproduce the measured quenched fraction as a function of satellite stellar mass. In particular, the quenching timescales that we infer steadily decrease with increasing satellite stellar mass, going from $\sim 1.6~{\rm Gyr}$ at $10^{10}~\msun$ to $\sim0.6~{\rm Gyr}$ at $10^{11}~\msun$. 

In general, the relatively short quenching timescale that we infer is consistent with previous studies at $z \sim 1$. For example, analyzing a sample of clusters from GCLASS, including some of the systems studied herein, \citet{Muzzin14} utilize the location of post-starburst galaxies within the cluster to infer a satellite quenching timescale of $\sim1~{\rm Gyr}$ for a sample of satellites with a median stellar mass of roughly ${\rm a~few} \times 10^{10}~\msun$. 
Likewise, using stellar population modeling to infer the rest-frame color evolution of satellites in $4$ clusters at $z \sim 1.5$, \citet{Foltz18} find a quenching timescale of $\tau_{\rm quench} \sim 1.1~{\rm Gyr}$ for satellites with $\mstar \gtrsim 10^{10.5}~\msun$. 
Finally, \citet{Balogh16} utilize a method similar to that employed in our analysis and allow for a quenching timescale that depends on stellar mass within a sample of GCLASS clusters at $z \sim 1$. At satellite stellar masses of $>10^{10}~\msun$, however, \citet{Balogh16} find a remarkably constant quenching timescale as a function of satellite mass ($\tau_{\rm quench} \sim 2~{\rm Gyr}$). 
While our estimates of the field quenched fraction are similar to those utilized by \citet{Balogh16}, the infall time distribution of our satellite population -- as inferred from the TNG simulations -- depends non-negligibly on satellite mass, such that lower-mass satellites are preferentially accreted earlier.
Quantitatively, we find the median difference in infall times to be about $0.4$ Gyr between galaxies with stellar masses of $10^{10}~\msun$ and $10^{11}~\msun$.
In contrast, \citet{Balogh16} adopt a model where the accretion history of their clusters depends only on host halo mass and not the mass of the satellite. 
In addition, the infall times adopted by \citet{Balogh16} are taken with respect to first infall onto any more massive halo (versus just the cluster halo, \citealt{McGee09}). 
These differences may account for the lack of mass dependence inferred in that work.

As a low-$z$ comparison, in Fig.~\ref{fig:tq_vs_mstar}, we show the quenching timescale inferred for the highest-mass clusters from the \citet{Wetzel13} sample (i.e.~$\mhalo=10^{14-15}~\msun$), which should roughly correspond to the descendants of our $z\sim1$ cluster sample.\footnote{While the typical GOGREEN cluster will evolve into a system with $\mhalo \sim 10^{15}~\msun$ at $z \sim 0$, our simulated cluster population will evolve into slightly less massive systems ($\mhalo \sim 10^{14.5}~\msun$ at $z \sim 0$).}
Scaling our results at $z \sim 1$ according to the evolution in the dynamical time  -- i.e.~$\tau_{\rm{quench}}(\mstar) \times (1+z)^{-1.5}$ -- we find good agreement between our inferred mass-dependent satellite quenching timescale and that from \citet{Wetzel13}. 
In \S\ref{subsec:phys} and \S\ref{subsec:preproc}, we further examine our quenching timescale constraints with an eye towards the potential physical mechanisms at play.


\section{Discussion}
\label{sec:Discussion}

\subsection{Internal Quenching after Infall}
\label{subsec:intq}

In contrast to some previous studies of satellite quenching \citep[e.g.][]{Balogh16}, a fundamental assumption of our fiducial model is that environmental and internal quenching mechanisms are separable, such that only environmental processes are at play once a galaxy becomes a satellite within the cluster halo. 
That is, we construct our model to account for the impact of internal quenching mechanisms by referencing the coeval field quenched fraction at the time of infall.
This, however, inherently assumes that environmental quenching mechanisms dominate within the cluster.
To test the validity of this assumption we adopt an alternative approach that allows internal quenching mechanisms to continue operating unabated after infall. We simulate this scenario by modifying our fiducial quenching model such that we classify galaxies as star forming or quenched at the redshift of observation ($z_{\rm{obs}}$) instead of at $z_{\rm{infall}}$, then determine the satellite quenching timescale (still relative to infall) needed to achieve the measured satellite quenched fraction as a function of stellar mass. 
Interestingly, we find that this approach yields very similar results to the scenario in which galaxies are classified at $z_{\mathrm{infall}}$, with the resulting satellite quenching timescale ($\tau_{\rm quench}$) as a function of satellite stellar mass consistent within $\pm 0.02~{\rm Gyr}$ for the two formulations of the quenching model.

The relative unimportance of internal quenching post infall for satellites at $z \sim 1$ is, in part, due to the short satellite quenching timescales at this epoch.
In addition, the role of internal mechanisms after infall is minimized by the mass-dependent efficiency of internal quenching (see Fig.~\ref{fig:CANDELS_fq_vs_mstar}) combined with the stellar mass dependence of the infall time distribution, such that more massive galaxies are more likely to be quenched internally but also typically become satellites later than their low-mass counterparts. 
In other words, given the mass dependence of typical infall times and given that the field quenched fraction as a function of cosmic time increases more slowly (rapidly) for low-mass (high-mass) galaxies, we find that the typical quenched fraction inferred at $z_{\rm obs}$ and $z_{\rm infall}$ are quite similar, thus yielding relatively similar results for the satellite quenching timescale.
Overall, the aforementioned modification to our fiducial model indicates that internal quenching mechanisms play at most a secondary role to the environmental quenching mechanism(s) operating within clusters at $z\sim1$.


\begin{figure}
 \centering
 \hspace*{-0.25in}
 \includegraphics[width=0.95\linewidth]{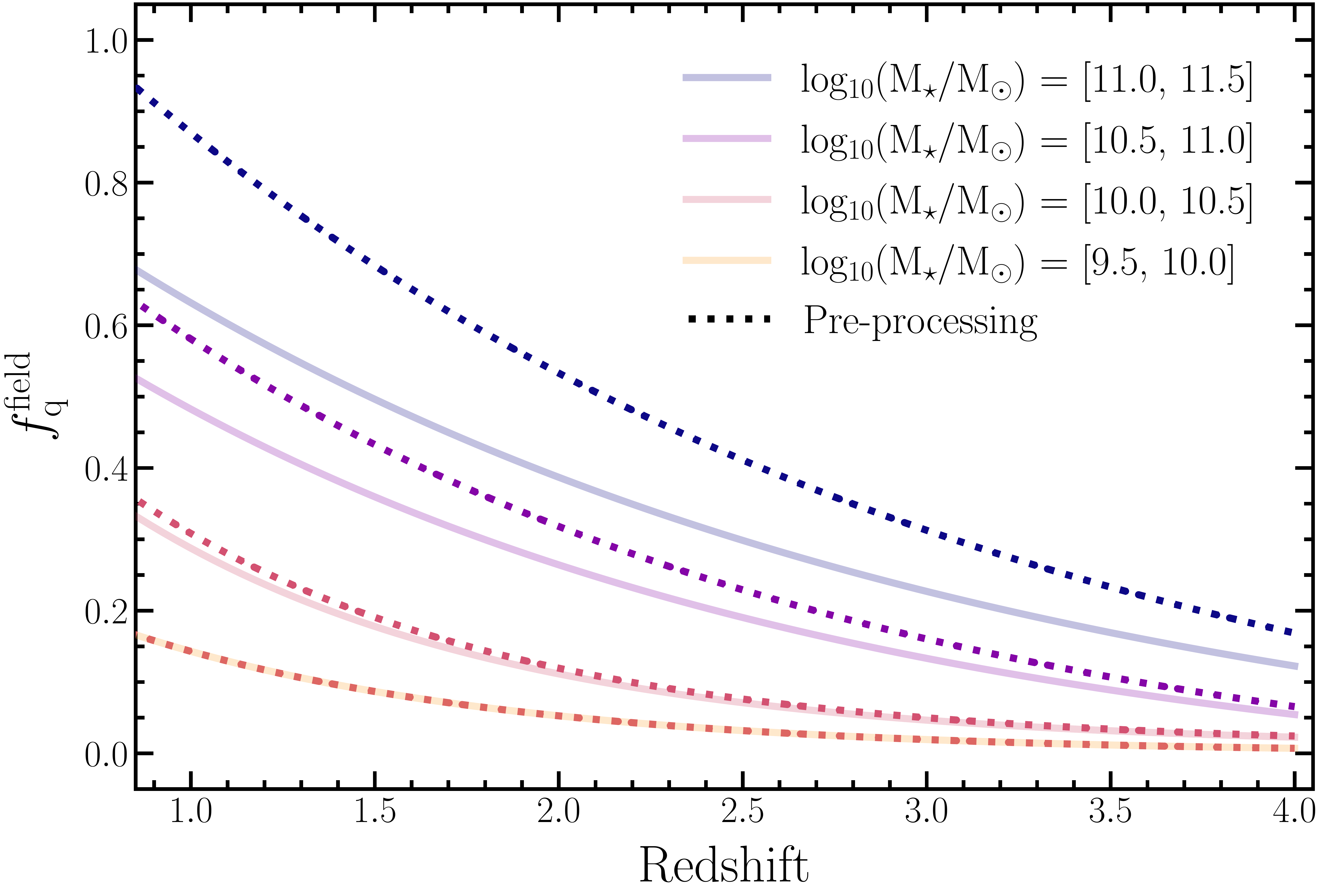}
 \caption{Field quenched fraction as a function of cosmic time and stellar mass. The faded lines represent our fits to the observed field quenched fraction from CANDELS (see Fig.~\ref{fig:CANDELS_fq_vs_mstar}). The dotted lines are the field quenched fraction results scaled to include the excess quenching due to additional satellite pre-processing in the infall regions of clusters. As discussed in \S\ref{subsec:preproc}, the scaling factor is derived from the measurement of the quenched fraction excess between the field and the infall region at $z \sim 1$ \citep{Werner22}.}
 \label{fig:field_quench_fraction_scaled}
\end{figure}


\subsection{Physical Processes Driving Satellite Quenching}
\label{subsec:phys}

The relatively long satellite quenching timescales inferred at low $z$ ($\tau_{\rm quench} \sim 4-7~{\rm Gyr}$) favor a slowly-acting quenching mechanism. Among the possible mechanisms, the long timescales for satellites at $\mstar \gtrsim 10^{9}~\msun$ strongly favor the starvation scenario by which satellites quench as a result of gas depletion in the absence of cosmological accretion following infall \citep{Wheeler14, Fham15, Fham16, Wetzel15}. 
As shown by \citet{Fham15}, the long satellite quenching timescales inferred for massive satellites in low-$z$ groups and clusters ($\mhalo \sim 10^{12-15}~\msun$) significantly exceed the molecular gas depletion timescales for similar systems at $0 < z < 2$ \citep{bigiel11, Saintonge11, Tacconi10, Tacconi13, Tacconi18, Freundlich19}. 
When factoring in the potential fuel supply associated with atomic gas, however, the dependence of $\tau_{\rm quench}$ on satellite stellar mass at $z \sim 0$ is shown to be in reasonably good agreement with the total cold gas (H$_{2}$ + H{\scriptsize I}) depletion timescale at $z \sim 0$ \citep{Fham15}. 

Measurements of the quenching timescale in lower-mass halos at $z \sim 1$ ($\mhalo \sim 10^{13-14}~\msun$) likewise yield timescales of $\sim 2-3~{\rm Gyr}$ at $\mstar \sim 10^{9.5-10.5}~\msun$ \citep[][but see also \citealt{Mok13, Mok14}]{Balogh16, Fossati17, Reeves21}. 
This exceeds the timescale upon which mechanisms like ram-pressure stripping are expected to act \citep{Tonnesen07, Bekki14} and also exceeds the molecular depletion timescale at the given mass scale and cosmic time \citep{Genzel10, Tacconi18}.
Similarly, while our fiducial model yields rapid quenching at high satellite masses, the inferred quenching timescale at lower masses ($\sim 10^{10}~\msun$) is longer than the molecular depletion timescale ($t_{\rm depl} \sim 0.5-1~{\rm Gyr}$) for field samples at $z \sim 1-2$. 
With that said, some measurements of CO-based molecular gas masses in $z > 1$ clusters do indicate that gas fractions (and depletion timescales) may be elevated in cluster populations \citep{Noble17, Noble19, Hiyashi18}. Other recent studies, however, find little variation in the molecular depletion timescale with environment \citep{Rudnick17, Williams22} or argue for depressed gas levels and thus shorter depletion timescales in high-density environments \citep{Alberts22}.


\begin{figure*}
 \centering
 \hspace*{-0.1in}
 \includegraphics[width=0.95\linewidth]{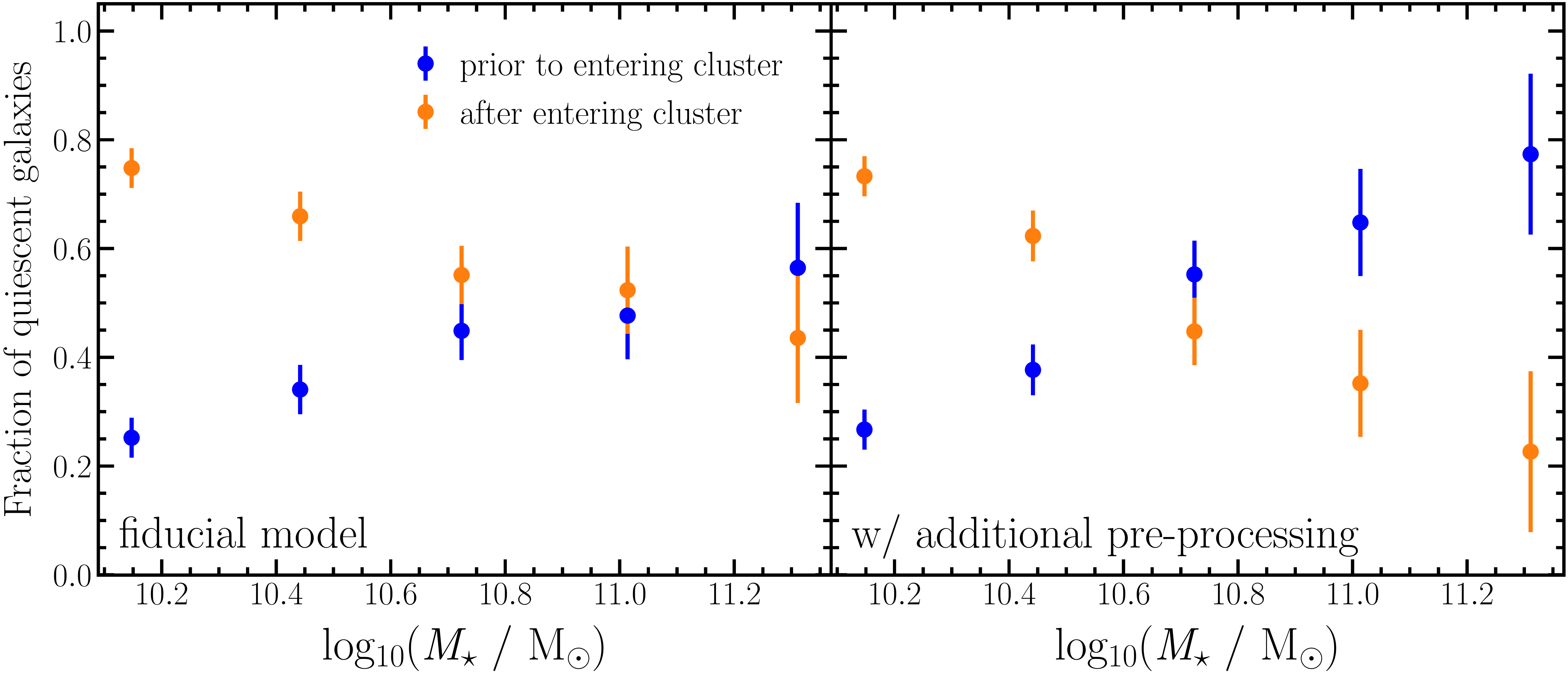}
  \caption{For the population of quiescent satellite galaxies in our model, we plot -- as a function of stellar mass -- the fraction of systems that were quenched prior to infall (blue points) versus quenched after infall (orange points) onto the cluster. The \emph{left} panel shows results for our fiducial quenching model, while the \emph{right} panel corresponds to results with additional pre-processing included (see \S\ref{subsec:preproc}). At the highest masses ($\mstar \gtrsim 10^{11}~\msun$), the majority of satellites are quenched prior to infall onto the cluster host halo, especially when accounting for pre-processing.}
 \label{fig:fq_vs_mstar_insideout}
\end{figure*}


Including atomic gas as a potential fuel for star formation, our quenching model yields satellite quenching timescales in closer agreement to the total cold gas (H$_{2}$ +  H{\scriptsize I}) depletion timescale at intermediate redshift. Given the typical infall time of our simulated sample, we include in Fig.~\ref{fig:tq_vs_mstar} the atomic + molecular depletion timescale as a function of stellar mass from the semi-empirical modeling of the gas reservoirs of galaxies as a function of cosmic time \citep{Popping15}. 
As found at $z \sim 0$, the relative agreement between the total cold gas depletion timescale and the satellite quenching timescale favors a scenario in which environmental quenching is driven by starvation. 
Moreover, similar to results at $z \sim 0$, where the satellite quenching timescale in groups and clusters shows little dependence on host halo mass for massive satellites \citep{Wetzel13}, current measurements of $\tau_{\rm quench}$ at $z \sim 1$ point towards a relative lack of variation in satellite quenching efficiency with host halo mass \citep[see \S\ref{subsec:tau};][]{Balogh16, Fossati17}. This further supports a picture in which satellite quenching is driven by starvation and follows a timescale dictated by the depletion of fuel for star formation following infall.

At high stellar masses, the cold gas (H$_{2}$ +  H{\scriptsize I}) depletion timescale does exceed the quenching timescale. However, it may be that the depletion timescales from \citet{Popping15} overestimate the atomic fraction in these systems -- as measurements of gas density in star-forming systems at intermediate redshift suggest a lower atomic component \citep[e.g.][]{Tacconi13} and some simulations predict a decrease in the atomic fraction in high-mass galaxies at $z > 1$ \citep{Dave17}. 
In addition, our model may underestimate the role of pre-processing that occurs prior to accretion, especially at high masses where increasing numbers of quenched ultra-massive galaxies have been identified in field surveys \citep[e.g.][]{Forrest20a, Forrest20b, Valentino20, McConachie21, Werner22}. As discussed in \S\ref{subsec:preproc}, including pre-processing within the infall regions surrounding our simulated clusters would lead to a corresponding lengthening of the satellite quenching timescale in Fig.~\ref{fig:tq_vs_mstar}, especially at $\mstar \gtrsim 10^{10.5}~\msun$. 
Another possibility is that complementary physical processes, such as ram-pressure stripping or feedback, are acting to decrease the reservoir of cold gas within satellites. 
Observations, both locally and at intermediate redshift ($z \lesssim 1$), find that stripping is clearly an active process in massive clusters \citep[e.g.][]{Poggianti17, Vulcani17, Boselli19, Moretti22}. 
Alternatively, stripping can also lead to increases in the surface density of star formation activity in satellite systems \citep{Merluzzi13, Vulcani18, Vulcani20}, which could contribute to expediting starvation via feedback \citep{McGee14}.

\subsection{Role of Pre-Processing}
\label{subsec:preproc}

Several studies of environmental quenching at low and intermediate $z$ find that ``pre-processing'' plays an important role in the build up of quiescent galaxies \citep[e.g.][]{McGee09, Cybulski14, Hou14, Just19, Pallero19, Sengupta22}. This occurs when a galaxy is subjected to environmental quenching as a consequence of becoming a satellite of a more massive galaxy prior to infall onto a group or cluster (or possibly via a filament, \citealt{Sarron19, GL22}). 
Our infalling satellite population is modeled using the ``field" quenched fraction from CANDELS (\S\ref{subsec:4.2}), such that our fiducial model includes some quenching due to pre-processing in lower-mass groups. This built-in level of pre-processing is most significant at lower satellites masses in our sample, where the fraction of satellite galaxies (relative to centrals) is greater. 

Recent studies have attempted to quantify the role of pre-processing through measurements of the quenched fraction excess (QFE, \citealt{VanDenBosch08}), which is also referred to in the literature as the conversion factor or quenching efficiency and defined as 
\begin{equation}
\label{eqn:QFE}
{\rm QFE}_{2-1} = \frac{f_{\rm q,2}-f_{\rm q,1}}{1-f_{\rm q,1}} \; ,
\end{equation}
where $f_{\rm q,2}$ is the fraction of quenched galaxies in a given environment (e.g.~the cluster regime) as compared to that in another environment (e.g.~the field or infall region surrounding a cluster, $f_{\rm q,1}$). 
In this context, a QFE of zero implies that there is no excess quenching between the two probed environments, while a QFE of one indicates that all star-forming galaxies in a given environment would be quenched were they to reside in the second (typically higher-density) environment.

\citet{Werner22} presents a relevant and recent study of pre-processing for satellites of GOGREEN clusters at $0.8 \lt z \lt 1.4$ by computing the QFE between coeval cluster, infall (\emph{inf}, $1 < R_{\rm proj}/R_{200} < 3$), and control (\emph{con}) field samples. They find that QFE$_{inf-con}$ strongly correlates with stellar mass, such that high-mass galaxies ($\mstar \sim 10^{11}~\msun$) that are star forming in the field are more likely to be quenched in the infall regions relative to their lower-mass ($\mstar \sim 10^{10}~\msun$) counterparts. 
To incorporate the impact of pre-processing in our quenching model, we scale our field quenched fraction as a function of redshift and stellar mass from Fig.~\ref{fig:CANDELS_fq_vs_mstar} by the aforementioned ${\rm QFE}_{inf-con}(\mstar)$ results from \citet{Werner22}. 
As shown in Figure~\ref{fig:field_quench_fraction_scaled}, this effectively augments the field quenched fraction of the most massive field galaxies (i.e.~$\mstar = 10^{11-11.5}~\msun$) as a function of redshift, such that a higher fraction of high-mass galaxies are quenched prior to infall. At low-masses, the level of pre-processing is significantly less, with the field quenched fraction largely unchanged relative to that utilized in our fiducial model. 
As illustrated in Figure \ref{fig:fq_vs_mstar_prep} and \ref{fig:fq_vs_z_prep} in Appendix \ref{sec:appendix}, the quenching model is specifically tuned to reproduce the observed quenched fraction as a function of stellar mass, however, it also reproduces the correlation between the quenched fraction and projected cluster-centric radius and redshift within the GOGREEN survey. 

As shown in Figure~\ref{fig:fq_vs_mstar_insideout}, including pre-processing increases the fraction of satellites that are quenched prior to infall onto the simulated clusters. 
This effect is most pronounced at higher stellar masses, with $\sim 65-80\%$ of simulated satellites quenched prior to infall at $\mstar > 10^{11}~\msun$ with the inclusion of pre-processing. 
In contrast to the results presented in Figure~8 from \citet{Werner22}, however, we do not find that $>90\%$ of ultra-massive ($>10^{11}~\msun$) galaxies are quenched prior to infall. In general, we find that the importance of pre-processing is likely weaker. %
In part, our results differ due to our more complete modeling of the accretion histories of satellite galaxies in our cluster sample. 
Comparing the quenched fractions of coeval populations via a measure of QFE partially ignores the evolution in those populations. Put simply, when compared to a sample of cluster members at $z \sim 1$, the coeval infall population does not represent the properties of the satellite population at the time of infall. Instead, a large fraction of the satellites in a cluster at $z \sim 1$ were accreted at $z \gtrsim 1.5-2$. 
Moreover, it is likely that our estimate of the quenched fraction for the ``pre-processed" population of infalling satellites is slightly overestimated. Studies of the QFE within groups and clusters as a function of cosmic time suggest that QFE (at fixed stellar mass) decreases with increasing redshift \citep{Lemaux19, Sarron21}. As such, by scaling our field quenched fractions by the ${\rm QFE}_{inf-con}$ at $z \sim 1$ from \citet{Werner22}, we likely overestimate the quenched fraction within infall regions at higher $z$. Similarly, a more complete analysis of the infall region would also factor in the contribution from quenched back-splash galaxies, which were quenched within the cluster but now reside within the infall regions \citep[e.g.][]{Balogh00, Gill05, Fham18}.  

\begin{figure*}
 \centering
 \hspace*{-0.1in}
 \includegraphics[width=0.65\textwidth]{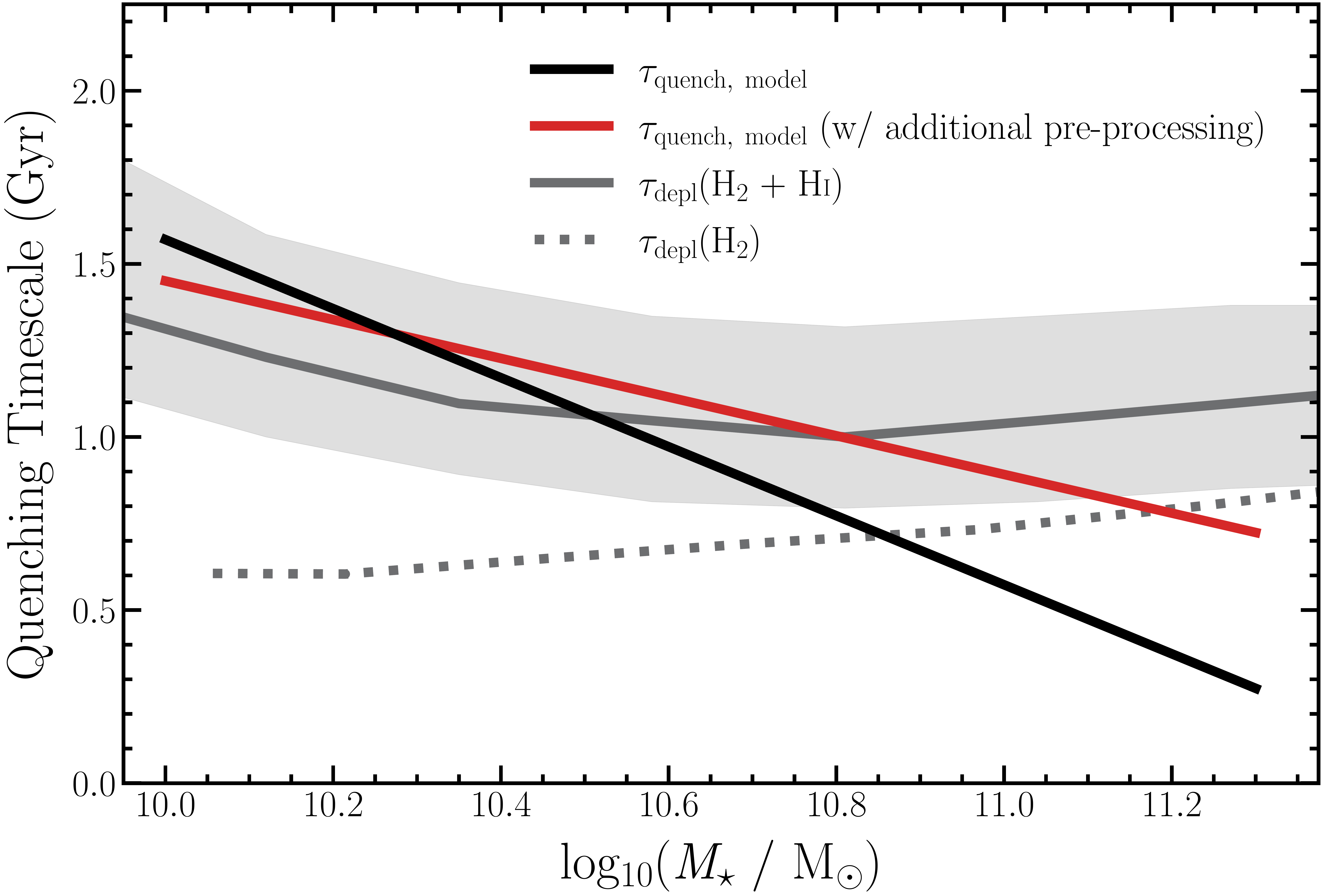}
 \caption{Quenching timescale versus stellar mass. The black solid line represents the quenching timescale results from our fiducial model, while the crimson line shows the results from our model including additional pre-processing (see \S\ref{subsec:preproc}). The solid grey line illustrates the empirically-derived cold gas (H{\scriptsize I} + H$_2$) depletion timescale from \citet{Popping15} at $z \sim 1.5$, with the grey shaded region corresponding to the variation in the depletion timescale over the redshift range $1 < z < 2$. Finally, the dotted grey line denotes the median molecular depletion timescale for our simulated infalling satellite population based on the scaling relations of \citet{Tacconi18}. Including additional pre-processing, we find a quenching timescale that is less strongly dependent on satellite stellar mass and is roughly consistent with the estimated cold gas (H$_{2}$ + H{\scriptsize I}) depletion timescale at $z \sim 1-2$.}
 \label{fig:tau_quench_all}
\end{figure*}

By accounting for pre-processing in our quenching model, we find that the best-fit quenching timescale is less strongly dependent upon stellar mass as shown in Fig.~\ref{fig:tau_quench_all}. 
At all masses, the inferred quenching timescale exceeds the typical depletion timescale for molecular gas. 
In Fig.~\ref{fig:tau_quench_all}, we illustrate the median molecular depletion timescale as a function of stellar mass for our simulated infalling satellite population based on the measured mass and redshift dependence of the depletion timescale for galaxies on the star-forming main sequence from \citet{Tacconi18}, adopting the relationship between star formation rate and stellar mass from \citet{Speagle14}. 
For comparison, we also include the empirically-derived H$_{2}$ + H{\scriptsize I} gas depletion timescale for galaxies at $z=1.5$ from \citet{Popping15}. The predicted cold gas depletion timescale depends on redshift at $1 < z < 2$, decreasing with increasing $z$ over the redshift range where a large fraction of our simulated satellite population is accreted. 
With pre-processing included in our model, the resulting satellite quenching timescale at $z \sim 1$ is in relatively good agreement with the cold gas (H$_{2}$ + H{\scriptsize I}) depletion timescale at intermediate redshift, similar to results at $z \sim 0$ \citep{Fham15} and consistent with starvation as the dominant mechanism for satellite quenching. 


\subsection{Impact of Stellar Mass Estimation}
\label{subsec:mstar_impact}

As discussed in \S\ref{subsec:3.2}, our fiducial model makes use of stellar masses from TNG that are defined to include the sum of all star+wind particles gravitationally bound to a given galaxy. A minor change would be to define stellar masses as the sum of all gravitationally bound star+wind particles within twice the stellar half-mass radius. We find that this change simply shifts the stellar masses lower by an average of $\sim0.1$ dex, but it does not significantly modify the results from the fiducial model. As shown in Fig.~\ref{fig:mstar_dist}, our fiducial model reproduces the relative distribution of satellite stellar masses for both the star-forming and quenched populations within GOGREEN.

Another aspect of our model is that it effectively defines crude star formation histories (SFHs) for the simulated satellites (e.g.~explicitly determining when particular systems quench); these SFHs may thereby differ from those within the TNG hydro-dynamical simulation, which are closely coupled to the stellar masses. 
Therefore, an alternative approach, which would more fully decouple our results from the prescriptions of baryonic physics utilized within TNG, is to define our simulated satellite stellar masses according to the assumption of a stellar mass-halo mass (SMHM) relation. We accomplish this using the \citet{Behroozi13} SMHM relation which estimates the stellar masses of galaxies using their peak halo mass and corresponding redshift. 
Compared to the TNG masses utilized in our fiducial model, the stellar masses inferred from the \citet{Behroozi13} SMHM relation are systematically less massive (by a few tenths of a dex). This bias towards lower masses is partially driven by a lack of ultra-massive galaxies ($>10^{11}~\msun$) predicted via abundance matching. 
Consequently, the observed distribution of satellite stellar masses from GOGREEN is not reproduced when assuming the \citet{Behroozi13} SMHM relation, in contrast to our fiducial model. 
However, when inferring stellar masses via abundance matching, we find that the resulting satellite quenching timescales -- $\tau_{\rm quench}(\mstar)$ -- are only slightly shorter (by $\sim0.1-0.2~{\rm Gyr}$) relative to those of our fiducial model.

\subsection{Success of Our Model}
\label{subsec:6.6}

Overall, our satellite quenching model reproduces many of the major observables from the GOGREEN survey -- the quenched fraction as a function of stellar mass (by construction), projected cluster-centric radius, and redshift. As a result, our model also reproduces the measured QFE as a function of stellar mass from \citet{vdB20}. Finally, our model likewise yields the observed stellar mass functions for both star-forming and quenched systems \citep{vdB20}. 
As shown in Figure~\ref{fig:mstar_dist}, our model reproduces the relative distribution of galaxy stellar masses for the quenched and star-forming populations in comparison to the corresponding observed counts from GOGREEN. With respect to the normalization of the resulting mass functions, our model underpredicts the total number of satellites due to our simulated clusters being biased towards lower halo masses (see Fig.~\ref{fig:m200_vs_z}). As discussed in \S\ref{subsec:3.1}, however, the distribution of infall times for our simulated satellites is weakly dependent on host halo mass (at $z > 1$ and $10^{14} < \mhalo/\msun < 10^{15}$), such that an increase in the number of satellites would not impact our measured satellite quenched fractions (i.e.~the results of the model).

While a quantitative comparison is beyond the scope of this work, the relatively short satellite quenching timescales (thus efficient environmental quenching) inferred by our modeling would yield older stellar ages and less extended SFHs for the GOGREEN cluster population relative to field galaxies of the same stellar mass. This is in agreement with recent results from \citet{Webb20}, which find that satellites within the GOGREEN clusters are typically $\sim0.3$~Gyr older than their field counterparts, with less extended SFHs. 
In addition, measurements of galaxy morphologies within the GOGREEN clusters find an excess of quiescent disks, particularly at low stellar masses \citep{Chan21}, which is also consistent with our results. Suppressing star formation via starvation will preferentially yield disky systems relative to processes such as mergers or harassment \citep[e.g.][]{Mastropietro05, Cortese07}. As found in the observations, within our model, the difference between the field and cluster morphologies should be most significant at lower satellite masses, where the environment plays a greater role in quenching (e.g.~see Fig.~\ref{fig:fq_vs_mstar_insideout}). 

Altogether, our model of satellite quenching is remarkably successful. In contrast, modern simulations of galaxy evolution tend to greatly overproduce the quenched satellite population at intermediate redshift, particularly at lower satellite masses (\citealt{Donnari21}; Kukstas et al.~in prep). This over-quenching problem is a long-standing one \citep[e.g.][]{Font08, Kimm09, Weinmann12, Hirschmann14, Wang14, Bahe17}, though progress has been made recently in reproducing observations of groups and clusters at $z \sim 0$ \citep[e.g.][]{DL19, Xie20, Donnari21}.

\section{Summary and Conclusions}
\label{sec:Conclusion}

Using simulated cluster and satellite populations from TNG, we model the quenching of satellite galaxies at $z > 1$ in comparison to observations from the GOGREEN survey.
The model includes one primary parameter, the satellite quenching timescale ($\tau_{\rm quench}$) that sets the time that a satellite remains star forming after infall onto the cluster. 
This timescale is tuned as a function of stellar mass to reproduce the observed satellite quenched fraction as a function of stellar mass. 
The main results from this modeling effort are as follows: 
\begin{enumerate}[leftmargin=0.25cm]
\item We measure the quenched fraction of GOGREEN cluster members as a function of stellar mass, projected cluster-centric radius, and redshift. We find that the satellite quenched fraction increases with stellar mass, decreases with projected radial cluster-centric separation, and remains relatively flat with redshift. \\
\item Our model reproduces the observed quenched fraction as a function of stellar mass (by construction), projected cluster-centric radius, and redshift as measured at $z \sim 1$ from the GOGREEN survey. In addition, our quenching model reproduces the relative galaxy stellar mass distribution (both in the field and in the cluster) as a function of galaxy type (star forming versus quenched). \\
\item In agreement with \citet{vdB20}, we find that satellite quenching is mass dependent at $z \sim 1$, in conflict with models that favor mass-independent environmental quenching \citep[e.g.][]{Peng10}. 
For our fiducial model, the quenching timescale depends on satellite stellar mass, such that galaxies at $\mstar = 10^{10}~\msun$ typically quench within $\sim 1.6~{\rm Gyr}$ following infall, while galaxies at $\mstar = 10^{11}~\msun$ quench much more rapidly (within $\sim 0.6~{\rm Gyr}$). 
Including pre-processing within the infall regions of clusters, the dependence of $\tau_{\rm quench}$ on satellite stellar mass weakens slightly, with satellites typically quenching on timescales of $\sim 1-1.5~{\rm Gyr}$ post infall, depending on mass. \\
\item In comparison to similar analyses at low redshift, we find that the satellite quenching timescale evolves roughly like the dynamical time ($\propto(1+z)^{-3/2}$), as noted by several previous studies \citep{Tinker10, Balogh16, Foltz18}. \\
\item When including pre-processing within the cluster infall regions, we find that the vast majority ($\sim 65-80\%$) of massive satellites ($>10^{11}~\msun$) in clusters are quenched at $z \sim 1$ clusters prior to infall. In contrast, the majority of lower-mass satellites ($\lesssim 10^{10.5}~\msun$) quenched within the cluster. \\
\item Our satellite quenching model yields quenching timescales that are longer than the observed molecular depletion timescales at intermediate redshift. Instead, the inferred quenching timescales are roughly consistent with the predicted total cold gas depletion timescale (H{\scriptsize I}+ H$_{2}$) at $1 < z < 2$. 
Similar to the results of modeling satellite populations in the local Universe, this may indicate that environmental quenching at $z > 1$ is primarily driven by starvation, where galaxies exhaust their fuel supply for star formation after being cut off from cosmological accretion. 

\end{enumerate}

\section*{acknowledgements} 
DCB thanks the LSSTC Data Science Fellowship Program, which is funded by LSSTC, NSF Cybertraining Grant $\#$1829740, the Brinson Foundation, and the Moore Foundation; participation in the program has greatly benefited this work.
FS acknowledges support by a CNES fellowship.
IPC acknowledge the financial support from the Spanish Ministry of Science and Innovation and the European Union - NextGenerationEU through the Recovery and Resilience Facility project ICTS-MRR-2021-03-CEFCA. 
GHR acknowledges the support of an ESO visiting
science fellowship. The authors would also like to acknowledge the support of the International Space Sciences Institute in Bern, who hosted a workshop that supported this publication.

This work was supported in part by NSF grants 
AST-1518257, 
AST-1716690, 
AST-1814159, 
and AST-1815475. 
Additional support was provided by NASA through grants 
AR-14289 
and 
AR-14310 
from the Space Telescope Science Institute, which is operated by the Association of Universities for Research in Astronomy, Inc., under NASA contract NAS 5-26555. 
GHR also acknowledges support provided via grant 
80NSSC19K0592  
issued through the NASA Astrophysics Data Analysis Program (ADAP).
MLB is supported by an NSERC Discovery Grant.
BV acknowledges suuport from the grant PRIN MIUR 2017 n.20173ML3WW\_001 (PI Cimatti) and  from the INAF main-stream funding program (PI Vulcani).
R.D. gratefully acknowledges support by the ANID BASAL projects ACE210002 and FB210003.

This research made extensive use of {\texttt{Astropy}},
a community-developed core Python package for Astronomy
\citep{Astropy13, Astropy18}.
Additionally, the Python packages {\texttt{NumPy}} \citep{numpy},
{\texttt{iPython}} \citep{iPython}, {\texttt{SciPy}} \citep{SciPy20}, {\texttt{Scikit-learn}} \citep{scikit-learn}, and
{\texttt{matplotlib}} \citep{matplotlib} were utilized for our data
analysis and presentation. 
In addition, this research has made use of NASA’s Astrophysics Data System Bibliographic Services.
Finally, this work makes use of observations taken by the CANDELS Multi-Cycle Treasury Program with the NASA/ESA HST, which is operated by the Association of Universities for Research in Astronomy, Inc., under NASA contract NAS5-26555.
\section*{Data availability} 
Data directly related to this publication and the figures within are available on request. The GOGREEN and GCLASS observational data used in this publication are accessible at the public data release website (\url{http://gogreensurvey.ca/data-releases/data-packages/gogreen-and-gclass-first-data-release/}), and NSF's NOIR-Lab (\url{https://datalab.noirlab.edu/gogreendr1/}). Likewise, the IllustrisTNG simulation data are publicly available and accessible at (\url{https://www.tng-project.org}). 
\bibliography{citations}

\appendix

\section{Quenched Fractions including Additional Pre-Processing}
\label{sec:appendix}

\begin{figure*}
 \centering
 \hspace*{-0.1in}
 \includegraphics[width=0.85\textwidth]{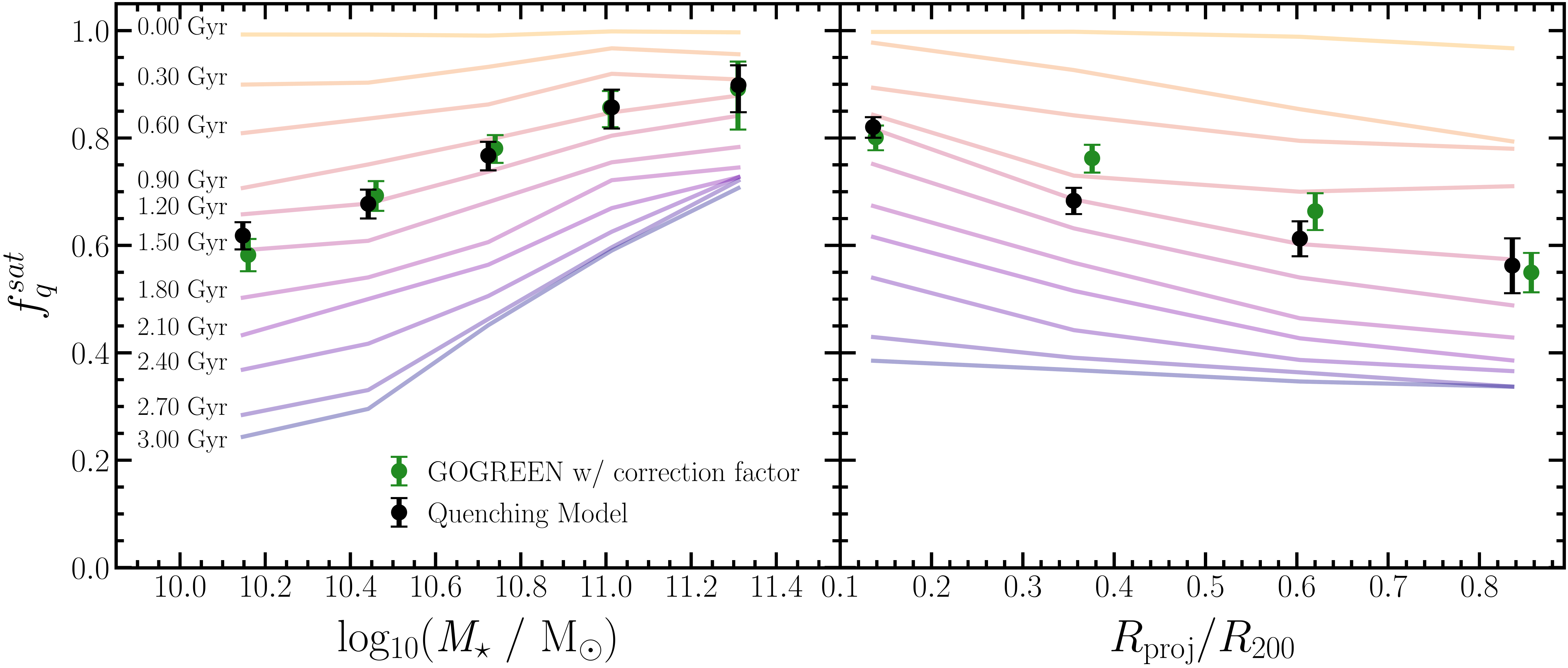}
 \caption{Satellite quenched fraction as a function of satellite stellar mass ({\it left}) and projected cluster-centric distance normalized by $\rtwo$ ({\it right}). Unlike Fig.~\ref{fig:fq_vs_mstar}, the results illustrated here are obtained using a modification to our fiducial quenching model designed to incorporate the effects of additional pre-processing. As before, the green circles illustrate the GOGREEN quenched fraction results with the membership correction factor applied. The colored translucent profiles in the background represent the TNG quenched fraction results using a constant quenching timescale ranging from 0 to 3 Gyr. The black circles represent the TNG results fit to the GOGREEN quenched fraction results. The observed quenched fraction as a function of stellar mass and cluster-centric radius are reproduced by a model assuming a mass-dependent quenching timescale, however, unlike the fiducial model it is clear that this modified model can reproduce both results by simply assuming a constant quenching timescale. All error bars represent 1-$\sigma$ binomial uncertainties.}
 \label{fig:fq_vs_mstar_prep}
\end{figure*}

\begin{figure*}
 \centering
 \hspace*{-0.1in}
 \includegraphics[width=0.5\textwidth]{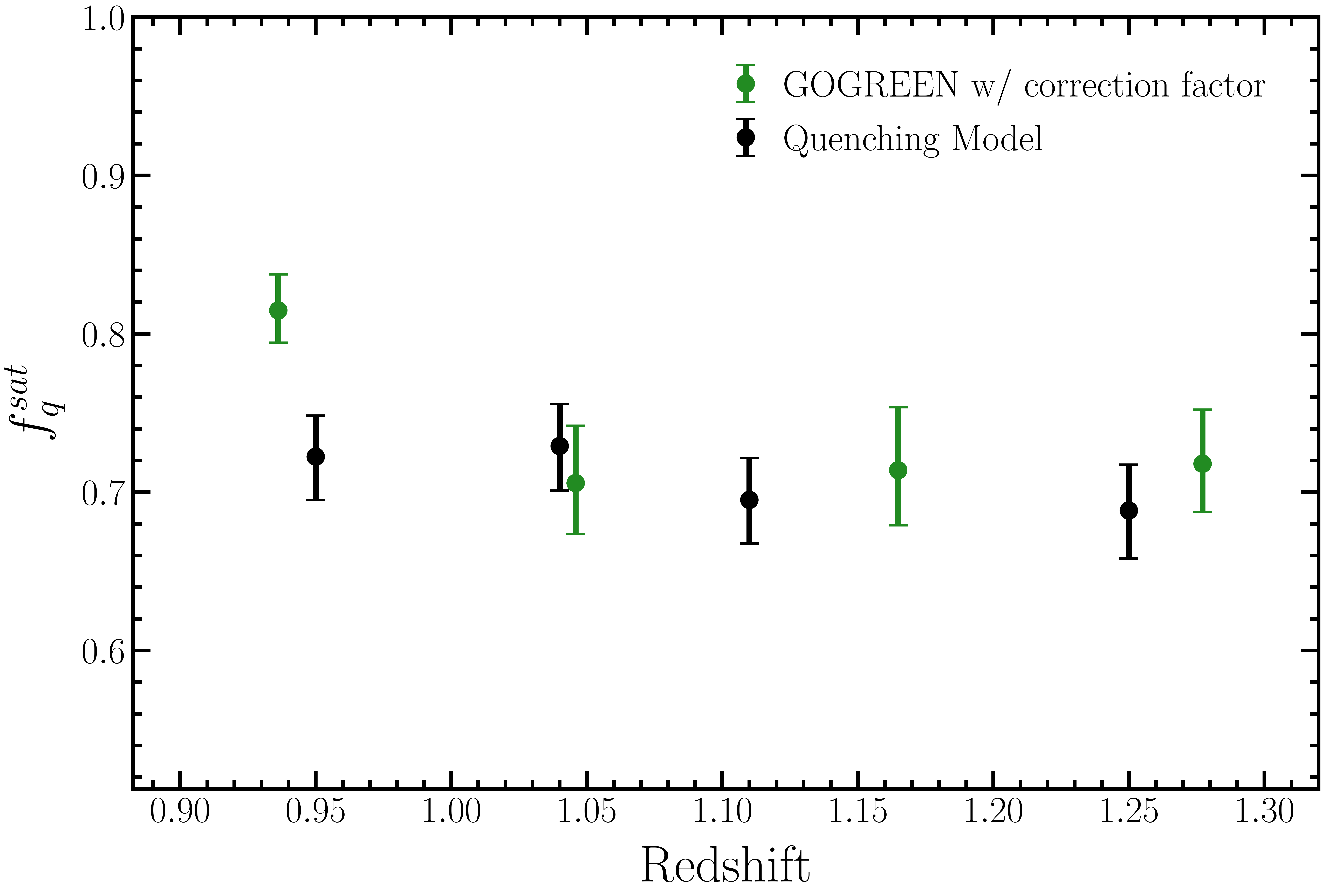}
 \caption{Satellite quenched fraction versus redshift. Unlike Fig.~\ref{fig:fq_vs_z}, the results illustrated here are obtained using a modification to our fiducial quenching model designed to incorporate the effects of additional pre-processing. The green circles represent the observed results with the membership correction applied. The black circles
shows the corresponding measurements for our modified fiducial model based on tuning $\tau_{\rm{quench}}$($\mstar$) to reproduce the observed satellite quenched fraction as a function of stellar mass. For both the observed and simulated samples, the uncertainties correspond to 1-$\sigma$ binomial errors. Our modified quenching model that incorporates additional pre-processing is also able to successfully reproduce the observed GOGREEN satellite quenched fraction as a function of stellar mass, projected cluster-centric radius, and redshift.}
 \label{fig:fq_vs_z_prep}
\end{figure*}

In Fig.~\ref{fig:fq_vs_mstar_prep} and Fig.~\ref{fig:fq_vs_z_prep}, we illustrate the results from our modified quenching model that incorporates additional pre-processing.


\section*{Affiliations}
\noindent {\it $\!\!^1$Department of Physics \& Astronomy, University of California, Irvine, 4129 Reines Hall, Irvine, CA 92697, USA \\
%
$\!\!^{2}$Department of Physics and Astronomy, University of Waterloo, Waterloo, ON N2L 3G1, Canada \\
$\!\!^{3}$Waterloo Centre for Astrophysics, University of Waterloo, Waterloo, ON N2L 3G1, Canada \\
$\!\!^4$School of Earth and Space Exploration, Arizona State University, Tempe, AZ 85281, USA \\
$\!\!^5$Departamento de Ingeniería Informática y Ciencias de la Computación, Universidad de Concepción, Chile \\
$\!\!^6$INAF - Osservatorio Astronomico di Trieste, via G.B. Tiepolo 11, 34143 Trieste, Italy \\
$\!\!^7$Departamento de Astronomía, Facultad de Ciencias Físicas y Matemáticas, Universidad de Concepción, Concepción, Chile \\
$\!\!^8$School of Physics and Astronomy, University of Birmingham, Birmingham, B15 2TT, UK \\
$\!\!^9$Department of Physics and Astronomy, York University, 4700 Keele St., Toronto, Ontario, M3J 1P3, Canada\\
$\!\!^{10}$Departamento de Ciencias Físicas, Universidad Andrés Bello, Fernández Concha 700, Las Condes, RM 7591538, Chile \\
$\!\!^{11}$Centro de Estudios de Física del Cosmos de Aragón (CEFCA), Plaza San Juan 1, 44001 Teruel, Spain \\
$\!\!^{12}$Department of Physics \& Astronomy, University of Kansas, 1251 Wescoe Hall Drive, Malott room 1082, Lawrence, KS 66045 \\
$\!\!^{13}$IRAP, Institut de Recherche en Astrophysique et Planétologie, Université de Toulouse, UPS-OMP, CNRS, CNES, 14 avenue E. Belin, F-31400 Toulouse, France \\
$\!\!^{14}$European Southern Observatory, Karl-Schwarzschild-Str. 2, 85748 Garching, Germany \\
$\!\!^{15}$INAF - Osservatorio astronomico di Padova, Vicolo Osservatorio 5, I-35122 Padova, Italy \\
$\!\!^{16}$Department of Physics and Astronomy, University of California, Riverside, 900 University Avenue, Riverside, CA 92521, USA \\
$\!\!^{17}$Steward Observatory and Department of Astronomy, 933 N. Cherry Ave, University of Arizona, Tucson, AZ, 85721 
}

\label{lastpage}
\end{document}